\documentclass[preprint,12pt]{elsarticle}
\usepackage[top=1.5in, bottom=1in, left=1in, right=1in]{geometry} 

\usepackage{amsmath, amssymb}
\usepackage{graphicx}  
\usepackage{dcolumn}   
\usepackage{bm}        
\usepackage[pdftex]{color}
\usepackage[utf8]{inputenc}
\usepackage{xcolor}
\usepackage{xspace}
\usepackage{hyperref}
\hypersetup{
    colorlinks=true,       
    linkcolor=blue,          
    citecolor=blue,        
    filecolor=blue,      
    urlcolor=blue           
}
\usepackage{slashed}
\usepackage{graphicx} 

\journal{arXiv}

\newcommand{\beq}{\begin{equation}}
\newcommand{\eeq}{\end{equation}}
\newcommand{\beqa}{\begin{eqnarray}}
\newcommand{\eeqa}{\end{eqnarray}}

\newcommand{\listing}[1]{\begin{center} \href{https://github.com/paboyle/Grid/blob/develop/#1}{#1}\end{center}}

\begin{document}

\begin{frontmatter}

\title{Multiple right hand side multigrid for domain wall fermions with
a multigrid preconditioned block conjugate gradient algorithm.}
\author{Peter Boyle}
\affiliation{organization={Physics Department},
            addressline={Brookhaven National Laboratory}, 
            city={Upton},
            postcode={11777}, 
            state={NY},
            country={USA}}

\begin{abstract}
We introduce a class of efficient multiple right-hand side multigrid algorithm for domain
wall fermions.
The simultaneous solution for a modest number of right hand sides concurrently allows
for a significant reduction in the time spent solving the coarse grid operator
in a multigrid preconditioner.
We use a 
preconditioned block conjuate gradient with a multigrid
preconditioner, giving additional algorithmic benefit from the multiple
right hand sides. There is also a very
significant additional to computation rate benefit to multiple right hand sides.
This both increases the arithmetic intensity in the
coarse space and increases the amount of work
being performed in each subroutine call, leading to excellent performance on modern
GPU architectures. Further, the software implementation makes use of 
vendor linear algebra routines (batched GEMM) that can make use of high throughput
tensor hardware on recent Nvidia, AMD and Intel GPUs.
The cost of the coarse space is made sub-dominant in this algorithm, and benchmarks from the Frontier supercomputer system show up to a factor of twenty speed up over the
standard red-black preconditioned conjugate gradient algorithm on a large system with physical quark masses.
\end{abstract}


\end{frontmatter}


\section{Introduction}

The \emph{critical slowing down} of fermion solver algorithms for the gauge
covariant Dirac equation is a key challenge for Lattice QCD. For a recent
review see ref.~\cite{Boyle:2024nlh}.
Multigrid algorithms have been developed that successfully 
address critical slowing down for Wilson and Wilson-clover fermions, 
\cite{Luscher:2007se,Brannick:2007ue,Brannick:2007cc,Clark:2008nh,Babich:2009pc,Osborn:2010mb,Frommer:2013kla,Frommer:2013fsa,Frommer:2011bad},
for calculation of valence propagators, while the application
to gauge configuration generation would certainly benefit from a reduced
setup cost. Multigrid has also been successful for twisted mass fermions\cite{Alexandrou:2016izb,Alexandrou:2018wiv,Bacchio:2019oiz}.
Significant challenges remain for staggered\cite{Weinberg:2017zlv,Brower:2018ymy,Ayyar:2022krp} and domain wall discretizations\cite{Cohen:2011ivh,Boyle:2014rwa,Yamaguchi:2016kop,Boyle:2021wcf,Brower:2020xmc}.

In the case of Wilson, clover and twisted mass Fermions the Dirac operator is one-hop and spectrally amenable to
invert with non-Hermitian Krylov solver algorithms, such as GCR and BiCGstab.                                             
However, Mobius and domain wall fermions have a spectrum that completely 
encloses the origin in the complex plane due to the large negative Wilson 
mass term and violates a \emph{half plane condition} required for effective
non-Hermitian solvers.\footnote{In the infinite volume limit, this spectrum is dense and a Krylov solver
must use a polynomial to approximate $1/z$ over this entire domain of the spectrum. 
Thus\cite{Boyle:2021wcf} there is an closed contour within the domain of the spectrum
over which one could integrate any possible Krylov polynomial 
and obtain zero, whereas the same integral of $1/z$ will yield $2\pi i$. They must therefore differ somewhere within the domain of the spectrum.} This results in very slow
convergence (of order the system size)\cite{Trefethen}. 
The consequence of this is that, to date, only four multigrid approaches to Domain Wall fermions have had any measure of success:
\begin{itemize}
    \item squaring the operator and coarsening a 2-hop (unpreconditioned squared matrix)\cite{Cohen:2011ivh}; and
    \item squaring the red-black preconditioned operator and coarsening a 4-hop matrix\cite{Boyle:2014rwa}; and
    \item restricting the action to the $c=0$ subspace of Mobius and coarsening the hermitian indefinite operator (with real spectrum)
$\Gamma_5 R_5 D_{dwf}$\cite{Yamaguchi:2016kop,Boyle:2021wcf}; and finally
\item coarsening both the Pauli Villars \emph{and} the Domain wall operator and solving the
combination $D(m=1)^\dagger D(m=m_{ud})$ in a non-Hermitian preconditioned system that successfully transforms the
spectrum to a half plane\cite{Brower:2020xmc}.
\end{itemize}

The dominant competing method is 
eigenvector deflation (with and without use of
a coarse compression scheme in local coherence Lanczos~\cite{Clark:2017wom}) and has
been widely used alternative for both staggered and domain wall fermions.
These are less complex than multigrid, but have a cost of $O(V^2)$ in terms of both computation and storage. They have been particularly attractive when combined with all-mode-averaging
\cite{Shintani:2014vja,Bali:2009hu}
and the all-to-all volume averaging approach\cite{Foley:2005ac}, since
efficient volume averaging in the low mode space
gives additional statistical advantages \emph{not} necessarily present using multigrid for certain calculations.

With domain wall fermions, deflating thousands of eigenvectors on typical lattices has
yielded a deflated solver that converges about twice as fast as HDCG.
However, for domain wall fermions with lattices of modest
size, such as the $48^3 \times 96 \times24$ physical
point ensemble used by RBC-UKQCD\cite{RBC:2014ntl}, a single ensemble data footprint
approaching several petabytes can be reached with 2000 eigenvectors per
configuration. In order to address finite volume effects
in lattice calculations of the hadronic vacuum polarisation
contribution to muon $g-2$ lattices with same parameters but sixteen times
larger physical volumes $64^3\times 128$ and $96^3\times 192$ have been
generated. 
These ensembles have the same parameters and are expected to have the same physical spectral \emph{density}, so 
we are required to deflate eigenvectors up to a 
fixed physical eigenvalue to obtain the corresponding solver acceleration.
Both the number of modes and the size of each mode must each grow
linearly in the volume. 
Thus both the cost of calculating, storing and reloading the
eigenvectors on such large lattices would be 256 times larger
and makes the approach unrealistic on these configurations
with current and near future machines and filesystems. An alternative algorithm
is required that either eliminates or radically reduces the
$O(V^2)$ cost of the calculations.

This paper presents such a solution in the form of a multiple right-hand side multigrid. 
We adopt the approach of \cite{Boyle:2014rwa} which was
successful and yielded order of magnitude gains for valence analysis on BlueGene computers, and likewise base our multigrid on the red-black preconditioned hermitian operator:
\beq
{\cal H} = (M_{ee} - M_{eo} M_{oo}^{-1} M_{oe})^\dagger (M_{ee} - M_{eo} M_{oo}^{-1} M_{oe}),\label{Eq:MpcDagMpc}
\eeq
where $M$ is the usual Mobius domain wall fermion Dirac operator\cite{Shamir:1993zy,Furman:1994ky,Brower:2004xi,Brower:2005qw,Brower:2012vk},
and the even and odd checkerboarding refers to partitioning the lattice sites
according to the parity of four dimensional lattice sites.
It improves upon the implementation of
the \emph{hierarchically deflated conjugate gradient}\cite{Boyle:2014rwa}, by reimplementing the algorithm
in a modern code base, Grid\cite{Boyle:2016lbp,Yamaguchi:2022feu}, 
\begin{center}
\href{https://github.com/paboyle/Grid/}{https://github.com/paboyle/Grid/},
\end{center}
giving portability to and high performance on Nvidia, AMD and Intel GPUs.
Finally, we extend the algorithm to a (multigrid) preconditioned block conjugate gradient for the first time.
This accelerates the convergence rate by an addition 25\% in outer iterations and was also computationally more efficient. 

It is intrinsic to the nature of multigrid algorithms that
the coarse grid corrections contain relatively little work.
This is precisely the opposite of the type of highly
parallel problem for which GPUs are designed. 
This problem leads to difficulty in optimising coarse operators
on GPUs. This paper formulates the problem
via coarse grid operator designed to work on multiple right-hand
sides concurrently, providing greater levels of parallelism. 
We will demonstrate that multiple right-hand side approach also provides many times more data reuse, improving the arithmetic intensity and floating point execution rate substantially. 
Strategies targeting matrix coefficient reuse across multiple vectors
have been used in domain wall fermions with a natural fifth dimension for quite some
\cite{Boyle:2009vp,Boyle:2012iy} and more recently adopted in the context of staggered
fermions \cite{Mukherjee:2014pye,Kaczmarek:2014mga}. This has been augmented with the observation that
with multiple right hand sides there can be algorithmic benefits that enhance the arithmetic gains by using
block conjugate gradient algorithms \cite{Clark:2017ekr,deForcrand:2018orx}.

Further, all elements of the coarse grid calculation 
can be formulated in terms of \emph{batched matrix-matrix} operations,
offloading a list of matrix multiplications to the GPUs via software
interfaces used widely in machine learning. These give access
to high throughput tensor floating hardware on top-end GPUs, 
exploiting hardware with increased peak performance.

The structure of this paper is as follows.
In section~\ref{sec:HDCG} we reprise the HDCG algorithm\cite{Boyle:2014rwa},
upon which this work is based, and extend the outer solver to a preconditioned block
conjugate gradient algorithm. We also discuss implementation details for multiple
right hand sides which makes the approach significantly more computationally efficient than
in the original work.
In section~\ref{sec:results} we present results for
physical quark mass inversions on a several volumes 
with inverse lattice spacing $a^{-1}=1.73$ GeV and up to $96^3\times 192$ volume, 
demonstrating excellent algorithmic volume scaling and exceeds 
by a factor of twenty the conventional red-black preconditioned conjugate gradient baseline.

\section{Multigrid and the HDCG algorithm}

\label{sec:HDCG}

We base our mrhs-HDCG algorithm on the 
ADEF2 two level preconditioned conjugate gradient of\cite{Tang}, figure~\ref{fig:pcg_general}.
We introduced two complementary preconditioners that are approximate the
inverse of the Dirac operator for the highest and lowest ends of the spectrum. 
So long as the composite preconditioned system is well conditioned the solver converges rapidly.
The low modes preconditioner captures much of the low mode degrees of freedom
by exploiting weak approximation
property\cite{alphaSA}, also known as local coherence\cite{Luscher:2007se}.

\begin{figure}
\begin{center}
Flexible ADEF2 Preconditioned Conjugate Gradient

{\small
\fbox{
\begin{minipage}{0.6\textwidth}
\begin{enumerate}
\item $x$ arbitrary
\item $M^{-1} := (P_R M_{IRS}({\cal H},\Lambda) + Q)$
\item $x_0 =  Q b + P_R x$
\item $r_0 = b - {\cal H} x_0$
\item $z_0 = M^{-1} r_0$ ; $p_0 = z_0$
\item for iteration $k\in 0,1,2\ldots$
\item \hspace{1cm}$d_k = {\cal H} p_k$
\item \hspace{1cm}$w_k = d_k - \sum\limits_{i=k-m_{max}}^{k-1} \frac{(d_k,{\cal H} w_i)}{(w_i,{\cal H} w_i)} w_i$
\item \hspace{1cm}$\alpha_k = (r_k,y_k)/(p_k,w_k)$
\item \hspace{1cm}$x_{k+1}= x_k + \alpha_k p_k$
\item \hspace{1cm}$r_{k+1}= r_k - \alpha_k w_k$
\item \hspace{1cm}$z_{k+1} = M^{-1} r_{k+1}$
\item \hspace{1cm}$\beta_k = (r_{k+1},z_{k+1})/(r_k,z_k)$
\item \hspace{1cm}$p_{k+1} = z_{k+1} + \beta_k p_k$
\item end for
\item return $x_{k+1}$
\end{enumerate}
\end{minipage}
}
}
\end{center}
\caption{\label{fig:pcg_general}
Flexible ADEF2 preconditioned conjugate gradient algorithm following Tang et al\cite{Tang} for solving a Hermitian system ${\cal H} x= b$, with
an additonal flexible search direction orthogonalisation (step 8).
We have implemented ``inexact preconditioned CG''\cite{GolubYe} and ``flexible CG'' \cite{Notay}
variants of the ADEF2 algorithm to address the variability in the
preconditioner when it is composed of Krylov processes. 
The matrix $Q$ is a coarse grid correction, and the matrix $M_{IRS}({\cal H},\Lambda)$ is a smoother in multigrid nomenclature. 
The Galerkin projector $P_R$ is defined in the text.
}
\end{figure}

The gauge freedom of our covariant Dirac operator is handled by generating vectors
$\phi_k$ in 
some way that lie within the space spanned by the lowest modes (the near null space of the operator). Once generated, these are subsequently chopped into many disjoint hyper-cuboidal sub-block vectors $\phi_k^b$ (or algebraically smooth wavelets)
with some blocking factor to form a basis for
our coarse grid. 
Coarse degrees of freedom are defined by taking the inner product with this wavelet basis, referred to
as a restriction operator $R=P^\dagger$ and a matching prolongation operator $P$, defining a projector to this subspace $S$ and it's complement $\bar{S}$.
\beq
P_S =  \sum_{k,b} |\phi^b_k\rangle \langle \phi^b_k | \equiv P P^\dagger \quad\quad ; \quad\quad P_{\bar{S}} = 1 - P_S
\eeq
A representation of the Dirac matrix may be formed within the subspace spanned by these $|\phi_k^b\rangle$
by calculating all non-zero matrix elements of the operator.
\beq
{\cal H}=
\left(
\begin{array}{cc}
{\cal H}_{\bar{S}\bar{S}} & {\cal H}_{S\bar{S}}\\
{\cal H}_{\bar{S}S} &{\cal H}_{SS}
\end{array}
\right)=
\left(
\begin{array}{cc}
P_{\bar{S}} {\cal H} P_{\bar{S}}  &  P_S {\cal H} P_{\bar{S}}\\
 P_{\bar{S}} {\cal H} P_S &   P_S {\cal H} P_S
\end{array}
\right)
\eeq
The coarse grid matrix is also known as the \emph{little Dirac operator},
\beq
h^{ab}_{jk} = \langle \phi^a_j| {\cal H} | \phi^b_k\rangle
\quad\quad ; \quad\quad
({\cal H}_{SS}) = h_{ij}^{ab} |\phi_i^a\rangle \langle \phi_j^b |. \label{eq:coarseop}
\eeq

$h$ inherits a sparse structure from ${\cal H}$ because well separated blocks do \emph{not} connect through ${\cal H}$.
We take the coarse grid correction in algorithm~\ref{fig:pcg_general},
\beq 
Q= \left( \begin{array}{cc}
0 & 0 \\
0 & {\cal H}_{SS}^{-1}
\end{array}\right).
\eeq

We can Schur decompose the matrix
\begin{eqnarray*}
{\cal H}= U D L = \left[ \begin{array}{cc}{\cal H}_{\bar{s}\bar{s}} & {\cal H}_{\bar{s}s} \\ {\cal H}_{s\bar{s}} & {\cal H}_{ss} \end{array} \right]
&=&
\left[ \begin{array}{cc} 1 & {\cal H}_{\bar{s} s}  {\cal H}_{ss}^{-1} \\ 0 & 1 \end{array} \right]
\left[ \begin{array}{cc} S & 0 \\ 0 & {\cal H}_{ss} \end{array} \right]
\left[ \begin{array}{cc} 1 & 0 \\ {\cal H}_{ss}^{-1} {\cal H}_{s \bar{s}} & 1 \end{array} \right]
\end{eqnarray*}
Note that
$P_L {\cal H} = \left[ \begin{array}{cc} S & 0 \\ 0 &0 \end{array} \right]$ yields the Schur complement $ S = {\cal H}_{\bar{s}\bar{s}} - {\cal H}_{\bar{s}s} {\cal H}^{-1}_{ss} {\cal H}_{s\bar{s}} $,
and that the diagonalisation $L$ and $U$ are related to Petrov-Galerkin oblique projectors:
\beq
P_L = P_{\bar S} U^{-1} =\left( \begin{array}{cc}
 1 & -{\cal H}_{\bar{S} S}  {\cal H}_{SS}^{-1}\\
 0 & 0
             \end{array} \right)
\quad\quad ; \quad\quad
P_R = L^{-1} P_{\bar{S}}  =
\left( \begin{array}{cc}
1 & 0 \\ -{\cal H}_{SS}^{-1} {\cal H}_{S \bar{S}} & 0
\end{array} \right).
\eeq

The coarse subspace inverse $Q = h^{-1} = {\cal H}_{SS}^{-1}$ is performed using deflated preconditioned Conjugate Gradients, and we precompute up to 1000 eigenpairs using block Lanczos\cite{Jang:2018cph}.
For multiple right hand sides, we extend the lattice for the vector space with an additional, fictitious
right hand side dimension. The matrix applied remains four dimensional, but is distributed across
the right hand side index saving memory bandwidth, in a manner similar to the application of four dimensional gauge link matrices in the 
the Wilson operator to multiple five dimensional fields in Domain Wall implementations\cite{Boyle:2009vp}.
Since the number of basis vectors is large, the memory reuse opportunity is particularly significant. In the software implementation, the coarse linear space is treated as
a five dimensional lattice field, similar to the Domain Wall fermion implementation in the Grid and BFM libraries\cite{Boyle:2009vp,Boyle:2016lbp,Yamaguchi:2022feu}.

HDCG took an ``infra-red shift'' smoother $M_{IRS}({\cal H},\Lambda)\sim({\cal H}+\Lambda)^{-1}$ which was approximated via a fixed number $N_{cg}=7$ iterations of conjugate gradients on the shifted matrix with a high-pass shift parameter
$\Lambda=2.0$. These are tunable algorithmic parameters.
The infra-red shift greatly stabilises the CG polynomial
and can be used to keep the polynomial sign definite over the range $[0,\infty)$ for even order polynomials, fixing issues with indefinite preconditioners
discussed in ref~\cite{OLEARY1991377}.
The separate right hand sides have the smoother implemented as distinct Krylov processes on each right hand side for largely software reasons.

\subsection{Role of block CG algorithms in multiple RHS multigrid}

In addition to arithmetic optimisation, the application of multigrid to multiple
right hand sides has an obvious potential benefit through the use
of block algorithms to share the searched Krylov spaces and give algorithmic acceleration.
However, the reason for super-linear convergence is obtained in BlockCG algorithms is subtle
and dependent on the spectrum of the matrices involved. The gains are typically gained at
larger iteration counts, and correspond to the removal of specific eigenvalues from the
residual polynomial\cite{Szyld} (so in some sense redundant with the effects targeted by deflation).

Unsurprisingly, we found no particular gain from applying BlockCGrQ, fig~\ref{fig:pbcgrq}, to our smoother: as the iteration count is very low so the subspaces are unlikely to mutually align. 

In the coarse grid solve, which we deflate using several hundred eigenvectors, 
we found marginal gain, at the order of 10\% our relatively cheap imprecise solve.
This is not enough gain to be worth pursuing since there is additional linear algebra and the
coarse solves are sub-dominant.
In fact, we find it simplest to treat the coarse solve as an enlarged five dimensional system with a 
single \emph{single} Krylov solve, albeit with a block diagonal structure. The number of conjugate gradient iterations is not significantly different compared to using independent conjugate gradient processes for each right hand side. However the linear
algebra overhead and software complexity are reduced. 

It is possible that the use of deflation
eliminates the low mode space where there is potential Krylov space co-linearity between right hand sides, and so deflation and block algorithms target largely the same
gain. Since it is possible and cheap to deflated the coarse space we find no
benefit to block algorithms in the coarse operator.

The final possible application of block solvers is to the outer Krylov solve.
Here we did find a meaningful gain, but the fact that a multigrid preconditioner is variable must
be addressed. 

\subsection{Preconditioned BlockCGrQ}

We generalise BlockCGrQ \cite{Dubrulle2001RetoolingTM,birk2015deflated}  (fig~\ref{fig:pbcgrq} left) to derive Preconditioned BlockCGrQ (fig~\ref{fig:pbcgrq} right). 
We use the approach taken in Saad Chapter 9\cite{saad} to derive the preconditioned conjugate gradient\cite{concus} from the
standard conjugate gradient algorithm\cite{HestenesSteifel}.
The key idea is that a standard conjugate gradient in the matrix $A$ with hermitian preconditioner
$M^{-1}$ and source $b$ is first written as a conjugate gradient in the matrix $M^{-1} A$ with source
$M^{-1} b$. 
The residual $z = M^{-1} A x - M^{-1} b = M^{-1} r$ and consistently the search directions
are updated in the preconditioner's guess for a solution given a residual in the original equations.
Finally, the  Euclidean inner products $\langle\rangle_E$ used in the rewritten CG taken as being in the $M$-norm $\langle\rangle_M$,
so that 
$$\langle z|z\rangle_M = \langle z M z \rangle_E= z^\dagger r,$$ 
and with a search direction $z = M^{-1} r$ in the rewritten CG produces Euclidean inner products with one power of the preconditioner removed.
Under this M-inner product $M^{-1} A$ is Hermitian because 
$
\langle x | M^{-1} A y \rangle_M
=  
\langle x | A y \rangle
=
\langle A x |  \rangle =
\langle M^{-1} A y |x \rangle_M
$.
With this substitution it is relatively easy to generalise BlockCGrQ to Preconditioned BlockCGrQ, figure~\ref{fig:pbcgrq}. We keep the Cholesky decomposition explict rather
than formulating in thin QR factorisation because the decomposition is performed
in the M-inner product, and the rotation of both the preconditioned
vectors $Z,Q$ and unpreconditioned vectors $z,q$ must be independently maintained.
The notation is intended to leave the relation to 
the BlockCGrQ for matrix $M^{-1} {\cal H}$ with M-inner product fairly evident.
BlockCGrQ algorithm was anticipated as possible to combine with preconditioning in ref~\cite{Dubrulle2001RetoolingTM}, but as far
as we know has not been discussed in the literature.

\begin{figure}
\begin{center}
{\small
\fbox{
\begin{minipage}{0.39\textwidth}
BlockCGrQ
\begin{enumerate}
\item ${\cal H} \in \mathbb{C}^{n\times n}$
\item $B$ source $\in \mathbb{C}^{n\times n_{rhs}}$
\item $X$ arbitrary $\in \mathbb{C}^{n\times n_{rhs}}$
\item $R_0 = B - {\cal H} X_0$
\item $Q_0 C_0 = R_0$
\item $D_0 = Q_0$
\item for iteration $k\in 0,1,2\ldots$
\item \hspace{1cm}$Z_{k} = {\cal H} D_{k}$
\item \hspace{1cm}$M_{k} = [D_{k}^\dagger Z_{k}]^{-1}$
\item \hspace{1cm}$X_{k+1}= X_{k} + D_{k} M_k$
\item \hspace{1cm}$Q_{k+1} S_{k+1}= Q_k - Z_k M_k$
\item \hspace{1cm}$D_{k+1} = Q_{k+1} + D_k S_{k+1}^\dagger$
\item \hspace{1cm}$C_{k+1} = S_{k+1} C_k$
\end{enumerate}
\end{minipage}
}
}
{\small
\fbox{
\begin{minipage}{0.5\textwidth}
Preconditioned BlockCGrQ
\begin{enumerate}
\item $z_0 = B - {\cal H} X_0$
\item $Z_0 = M^{-1} z_0$
\item $C_0^\dagger C_0 = Z^\dagger_0 \cdot z_0 = \langle Z_0| Z_0\rangle_M$
\item $q_0 = z_0 C^{-1}_0 \quad ; \quad Q_0 = Z_0 C^{-1}_0$
\item $D_0 = Q_0$
\item for iteration $k\in 0,1,2\ldots$
\item \hspace{1cm}$z_{k} = {\cal H} D_{k}$
\item \hspace{1cm}$Z_{k} = {M^{-1}} z_{k}$
\item \hspace{1cm}$N_{k} = [D_{k}^\dagger \cdot z_{k}]^{-1} = \langle D_{k} | Z_{k}\rangle^{-1}_M$
\item \hspace{1cm}$X_{k+1}= X_{k} + D_{k} N_k$
\item \hspace{1cm}$t_k =  q_k - z_k N_k \quad ; \quad T_k =  Q_k - Z_k N_k$
\item \hspace{1cm}$S_{k}^\dagger S_{k+1} = T^\dagger_k\cdot t_k = \langle T | T \rangle_M$
\item \hspace{1cm}$q_{k+1} = t_k S_{k}^{-1} \quad ; \quad Q_{k+1} = T_k S_{k}^{-1} $
\item \hspace{1cm}$D_{k+1} = Q_{k+1} + D_k S_{k}^\dagger$
\item \hspace{1cm}$C_{k+1} = S_{k} C_k$
\end{enumerate}
\end{minipage}
}
}

\end{center}
\caption{\label{fig:pbcgrq}
We generalise BlockCGrQ \cite{Dubrulle2001RetoolingTM,birk2015deflated}  (left) to derive Preconditioned BlockCGrQ (right). 
}
\end{figure}

\subsection{Multigrid preconditioner variability}

\label{sec:static}
We obtained a significant gain from the Preconditioned BlockCGrQ 
as the outer solver, after addressing the problem of multigrid preconditioner 
variability. In the single right hand side HDCG, we addressed this by introducing a flexible search direction orthogonalisation to ADEF2, but
given the linear algebra is proportional to the square of the
number of right hand sides the introduction of
orthogonalisation to the recent search directions in preconditioned
BlockCGrQ was less attractive than the option of replacing data
dependent Krylov processes in the preconditioner with a stationary
preconditioner. We note that retaining several times the number of right hand sides 
prior search directions would enable A-orthogonalisation to be performed even in the block case to 
establish a flexible preconditioned block CG, but at the expense of considerable storage.

It proved relatively simple to replace the typical polynomial that
is produced by the Conjugate Gradient algorithm at $n$ iterations  in both
the coarse solver and the smoother by a similar order Chebyshev approximation 
to $\frac{1}{x}$ that is tuned to approximate over the same range.
We simply replaced the 7 iteration smoother $M_{IRS}({\cal H},\Lambda)$ with high pass parameter $2.0$ with a 8th order Chebyshev approximation to $\frac{1}{x}$ over eigenspan $[lo,hi] = [2.0,92.0]$ having obtained the highest
eigenvalue of our matrix with a cheap application of the power method.

Similarly, we replace the deflated conjugate gradient method, solving to
$4\times10^{-2}$ residual from Conjugate Gradients in around 120 iterations,
with a two step deflated/Chebyshev coarse solve of the same order.
The source $b$ is split into pieces parallel and orthogonal to the deflation space consisting of the $n_{ev}$ lowest eigenvectors,
\begin{eqnarray}
b &=& b^\parallel + b^\perp,\\
b^\parallel &=&\left(\sum\limits_{i=1}^{n_{ev}} |i\rangle\langle i|\right) b,  \\
b^\perp &=& \left(1-\sum\limits_{i=1}^{n_{ev}} |i\rangle\langle i|\right) b, \\
Q b^\parallel &= &
\left(\sum\limits_{i=1}^{n_{ev}}\frac{|i\rangle\langle i|}{\lambda_i}\right) b.
\end{eqnarray}
The solve is then composed of a deflation inverse on $b^\parallel$ and
a Chebyshev polynomial approximation $\mathbb{P}^{\rm Cheby}$ to $\frac{1}{x}$ on the perpendicular subspace.
Since the coarse operator is Hermitian and diagonalisable the stationary
coarse solver is composed of two regimes in the eigenspace
\begin{eqnarray}
Q^{Cheby+defl}& = &
{\rm diag} (\frac{1}{\lambda_{min}},\ldots, \frac{1}{\lambda_{n_{ev}}},
\mathbb{P}(^{\rm Cheby}\lambda_{n_{ev}+1}), \ldots,
\mathbb{P}^{\rm Cheby}(\lambda_{max}))\\
Q^{Cheby+defl} b &=& {\rm Defl}(b^\parallel) + {\mathbb P}^{\rm Cheby}(b^\perp).
\end{eqnarray}
The range of the Chebyshev approximation was selected to exceed the upper end of the spectrum (40.0)
and come within a factor of ten of the highest eigenvalue in the deflation basis (0.02).
The tuned parameters of the Chebyshev are thus, $$
[ lo, hi, order] = [0.02, 40, 120],
$$
with coefficients chose according to the usual Chebyshev approximation obtained by determining coefficients
at the Chebyshev interpolation points from the target function $\frac{1}{x}$.
After tuning, the deflation
and polynomial approximation actually achieved a slightly reduced residual compared
to the Conjugate Gradient approach but brings the advantage the polynomial is stationary.

Further because linear algebra and reductions are reduced compared to conjugate
gradients, the application of the stationary multigrid preconditioner is actually faster while
a reduction of the outer iteration count by around 25\% is obtained in the preconditioned 
block algorithm. This ultimately 
leading to our fastest multiple right hand side multigrid solver implementation in the results
section~\ref{sec:results}.

\subsection{Basis setup}

\label{sec:subspace}

Since eigenvalues of the fine operator are gauge invariant but eigenvectors are covariant, the obvious way to generate these vectors and 
so ``define'' a local averaging in a block is by
using the covariant Dirac operator itself to spectrally filter out 
the near null space of the operator.

Lattice QCD multigrid typically generates
basis vectors $\phi_k(x)$ lying in the near null space 
by repeated application of an approximate inverse of
the Dirac matrix to a random vector. The first pass multigrid solver 
can be used to improve the subspace during a setup phase. 
Once obtained, these vectors are obtained, they are restricted to disjoint blocks $b$ and orthonormalised. The span of these block vectors  $\phi_k^b(x)$ is much larger than the span of the $\phi_k(x)$, and due to local coherence
includes many low eigenmodes as a subspace. 
The sharp edges between blocks, however, introduce high mode spectral leakage when a
coarse grid correction is applied. The composite preconditioner performs best if fine-grid smoothing is included.

We consider two types of set up in this
paper: a) Low pass filtering schemes and b) Lanczos derived eigenvectors of the fine operator.

\subsubsection{Filtering subspace definition}
Filtering schemes are a heuristic approach
applying some form of low pass filter of
the fine matrix to a given input noise
vector $\eta_k$. The commonly used Wilson multigrid
approach of inverse iteration is one
form of this class of set up scheme,
and if run to exact convergence, $n$ steps of inverse iteration produces a filter function ${\cal F}_n(\lambda) = \frac{1}{\lambda^n}$ applied to whatever the spectral
decomposition of the initial noise vector is to obtain $\phi_k = {\cal F}_n({\cal H}) \eta_k$.

With the Hermitian positive case, as considered in this paper, the reality of the spectrum is a significant advantage. Various forms of approximation of functions
of the matrix may be used knowing the spectrum lies on the positive real line.
For example, in the HDCG
paper a multishift conjugate
gradient solver was used on 
rational function to obtain a 4th order
low pass filter of the form,
\beq
{\cal F}_{rational}(\lambda) = \frac{1}{
(\lambda + \Lambda)
(\lambda + \Lambda+\epsilon)
(\lambda + \Lambda+2\epsilon)
(\lambda + \Lambda+3\epsilon)
},
\eeq
using a \emph{single} multishift solve
with low pass parameter $\Lambda=0.0003$ and a convenience parameter $\epsilon=\Lambda/3$ to ensure a partial fraction expansion can be performed so that
the problem is solvable as a multishift inversion.

In this paper we instead adopt some of the
Chebyshev filtering and approximation techniques introduced in ref.
\cite{Boyle:2021wcf}, to use Chebyshev derived low
pass filters. This was inspired by
the use of Chebyshev filters in polynomial preconditioned
Lanczos algorithm\cite{RudyPhD,Shintani:2014vja,doi:10.1137/0613025}.
In ref.~\cite{Boyle:2021wcf} the fact that Chebyshev
polynomials are recursively determined
was used to produce multiple linearly independent filter functions for minimal
cost. The aim of ref.~\cite{Boyle:2021wcf} was to obtain
sufficient set up speed that a gain in gauge
evolution was possible, whereas here with a focus on multiple right hand sides and valence analysis we are less concerned
with setup speed than with ultimate solver gain. We have therefore not attempted to optimise the time to generate a Chebyshev based approach making use of the recursive nature.

The filter scheme adopted was as follows, taking the n-th Chebyshev polynomial as $T_n(x) = \cos( n \cos^{-1}(x))$ and upper/lower eigenvalue parameters $b=92.0$ and $a_1=0.001$ and $a_2=0.02$:
$$
\phi_k = F_{cheby}({\cal H})\eta_k = T_{600}(2\frac{({\cal H}-a_2)}{b-a_2}+a_2) T_{2500}
(2\frac{({\cal H}-a_1)}{b-a_1}+a_1) \eta_k.
$$
There is considerable scope for cost reduction in this class of filtering
scheme, for example applying further filters to previously generated
vectors or using the recursive nature of Chebyshev polynomials\cite{Boyle:2021wcf}. 
This has not been investigated in detail as the cost of the
solver \emph{after} subspace generation has been the focus of this work.
The Chebyshev setup scheme above is substantially faster than running the full Lanczos and is already sufficient to obtain
a fast multigrid solver.

\subsubsection{Eigenvector based subspace definition}

The second scheme considered was partly of academic interest. We are not aware of systematic study of the optimal spectral
content in  multigrid basis generation, which is mostly heuristic and
empirical.
We therefore, on the $48^3\times 96$ test system only, computed the 200 lowest modes
of the fine Dirac operator and studied 
how subspace vectors with different spectral
content behave. These we call ``eigenvector'' based setup. 
We considered the cases of: 
\begin{enumerate}
\item taking the lowest $n_{basis}$ modes,
\item taking $n_{basis}$ but taking every $m-th$ mode for $m=2,3$, and finally
\item taking $n_{basis}$ sums of $m$ consecutive modes.
\end{enumerate}

We will demonstrate in section~\ref{sec:results} that 
the best deflation effect is obtained with the $n_{basis}$ lowest modes, but that
the multigrid speed is still fairly insensitive to the choice at the ten or twenty percent level. This is in stark contrast to eigenvector deflation where omission of a single mode will
lose deflation efficacy.

\subsection{Software implementation and multiple right hand sides}

There are relatively new, convenient and well optimised
machine learning targeted ``batched BLAS GEMM'' routines
for performing a list of many general matrix-matrix multiplications.
Software interfaces are available under CUDA, HIP, and SYCL interfaces easing portability across
modern GPUs. 
By using these interfaces, access is given to tensor multiplication hardware with
significantly increased peak and actual floating point throughput.
The performance is good because the matrix ranks are significantly larger than arises with $N_c = 3$ fine grid operations.
It is wise for Lattice software to take the opportunity to use AI and 
machine learning targeted hardware when appropriate.
It is relatively easy to obtain 20 to 40 Tflop/s per node in double precision on the Frontier supercomputer (for example), depending on matrix dimensions.

A GPL licensed listing for a portable interface to the relevant cublas, hipblas and OneMKL interfaces for GPUs
and a multithreaded call to the Eigen Library for CPU's is given in the Grid Library:
\listing{Grid/algorithms/blas/BatchedBlas.h}

The coarse operator and multigrid implementation in Grid has therefore been updated to support application
to multiple right hand sides concurrently because we can apply the operator to several right-hand sides
with very little increase in memory traffic and cost.  

When the number of basis vectors is large the volume of data in matrix elements $A_{ij}$ is much greater than the volume
of data in a coarse vector and the cost of applying the coarse operator is dominated by the cost of fetching these coefficient matrices.

\subsubsection{Transfer between grids}

Migration between coarse and fine grids is performed using batched GEMM
subroutines, introducing a `MultiRHSBlockProjector' object,
\listing{algorithms/deflation/MultiRHSBlockProject.h}

Data is laid out internally contiguous arrays.
The large basis vectors are imported to the object and stored in BLAS ready layout.
The course and fine vectors are imported into preallocated device arrays, calculation performed,
and then results exported on each call. Since the number of right hand sides is substantially lower
than the number of basis vectors the import and export overhead is relatively small compared
to the total amount of work.
Since the batched BLAS routines require pointer array arguments for the base addresses
of each matrix in a list, the copy-in to a preallocated array avoids allocating new
device vectors and calculating new pointer arrays upon each call, and is a reasonable
compromise as such allocations can be expensive compared to usage of GPU device resident
data.
The block project operation from fine degrees
of freedom $F$ to coarse degrees of freedom $C$ 
with a basis definition $B$ is performed as a coarse volume batched GEMM operation, with one matrix multiply for each coarse
lattice site $x_c$,
\beq
\forall_{x^c} \quad : \quad
C_{N_{basis} \times N_{rhs} }(x^c) = 
B^\dagger_{N_{basis}\times V_{block} }(x^c) \times  F_{V_{block} \times N_{rhs}(x^c)}
\eeq
while the block promote operation from coarse to fine degrees of freedom
is 
\beq
\forall_{x^c} \quad : \quad
F_{V_{block} \times N_{rhs}(x^c)}
= 
B_{V_{block} \times N_{basis} }(x^c)
C_{N_{basis} \times N_{rhs} }(x^c) 
\eeq

\subsection{Coarse operator}

Coarsening an operator requires calculating the non-zero matrix elements Eq.~\ref{eq:coarseop} of the
fine grid operator. 
The Grid software package\cite{Boyle:2016lbp,Yamaguchi:2022feu}
has been updated to support coarsening 1-hop, 2-hop and 4-hop operators, connecting via a
a general coarse matrix stencil ranging from displacements $(-1,-1,-1,-1)$ to $(1,1,1,1)$ for up to $3^3=81$ points in the stencil. This calculation accommodates 4 hop fine grid operators under the assumption that the blocking
size equals or exceeds $4^4$, so that the coarse grid representation does not require a shift by more than
one coarse grid site in any given direction.

The listing of the coarse operator, with the implementation of the batched BLAS optimised coarsening and the batched BLAS coarse operator
application is available in the Grid library:
\listing{Grid/algorithms/multigrid/GeneralCoarsenedMatrixMultiRHS.h}

The geometry of the coarse operator is specified by creating a list of all the relative shifts for which
there is an expected non-zero matrix element. For each basis vector $\phi_k$ and the relative shift $\delta^p$ for each known point $p$ in the expected stencil of non-zero connections, there is a sub-block by sub-block phase applied,
\beq| \tilde\phi^p_k \rangle =  \sum_b e^{i b \cdot q^p}| \phi_k^b\rangle,\eeq
with the `momentum' choice $q^p_\mu = 2\pi \delta^p_\mu /L_\mu$.
Then matrix elements are taken with blockwise inner products,
\beq
\label{Eq:coarsen}
\langle \tilde\phi_{j}^{p,b} | M | \tilde \phi_k^p \rangle = 
\sum_{b^\prime}\langle \phi_{j}^{b} | M | \phi_k^{b^\prime} \rangle  e^{i (b-b^\prime)\cdot q^p}
=
\sum_{p^\prime\in {\rm stencil}}
\langle \phi_{k^\prime}^{b} | M | \phi_k^{b+\delta^{p^\prime}} \rangle  e^{i \delta^{p^\prime}\cdot q^p} 
= 
\langle \phi_{k^\prime}^{b} | M | \phi_k^{b+\delta^{p^\prime}} \rangle  \alpha_{p,p^\prime}.
\eeq
The inner products in Eq.~\ref{Eq:coarsen} can be computed efficiently using the batched block projector.
With this choice we obtain an invertible phase matrix $\alpha_{p,p^\prime}= e^{i \delta^{p^\prime}\cdot q^p} $ that allows to
resolve the $N_{\rm stencil} \times N_{\rm basis} ^2$ matrix elements with $N_{\rm stencil} \times N_{\rm basis}$
fine matrix multiplies as,
\beq
\langle \phi_{k^\prime}^{b} | M | \phi_k^{b+\delta^{p^\prime}} \rangle 
= (\alpha^{-1})_{p^\prime,p} \langle \tilde\phi_{j}^{p,b} | M | \tilde \phi_k^p \rangle
.\eeq

{\bf Applying the coarse operator:}
The `MultiGeneralCoarsenedMatrix' object
provides a coarse operator itself is implemented
using a ``ghost zone'' halo exchange, in the Grid ``PaddedCell'' object,
\listing{Grid/lattice/PaddedCell.h}

The exchange is performed serially over dimensions. So that a larger local lattice is first build
containing the ghost zone in the x-direction first. This extended object then exchanges boundaries, including the
existing x-ghost zone corner, in the y-direction and so on. Thus the corner terms are communicated for the 81 point
stencil while still having only nearest-neighbour communication via this sequential scheme.
The communication cost therefore does not particularly grow with the stencil size.

The arithmetic for a single point $p$ in the stencil of the coarse operator with
displacement $\delta_p$ can be posed as a batched GEMM call with batch size the coarse grid local volume,
\beq
\forall_{x^c} \quad : \quad
C_{N_{basis} \times N_{rhs}}(x^c)
=
C_{N_{basis} \times N_{rhs}}(x^c) +
A^p_{N_{basis} \times N_{basis}}(x^c) \times 
B_{N_{basis} \times N_{rhs}}(x^c+\delta_p).
\eeq
The batched GEMM routines accept a pointer list
for the matrices $A$, $B$ and $C$.
This allows the indirect neighbour addressing to be
performed within the batched GEMM routine.
The input fermion vector $B$ has dimension of the padded
grid volume that includes boundary zones, 
while the result vector has support only on the properly
owned node-local volume. The pointers are precomputed in the `MultiGeneralCoarsenedMatrix' operator. 
The input coarse fermion vectors are imported and exported at minimal cost
to preallocated arrays and the coarse matrix coefficients
are stored entirely internal to this object.
If the number of right hand sides is significantly less than the number of basis vectors, the overhead of
import/export remains small.
The implementation is generic and
can coarsen one, two and four hop stencil fine operators
of any fermion type.

The (repeated) deflation of a source is necessary to accelerate the
coarse operator solution in the inner iteration.
This was optimised for the multiple right hand side case, again by
use of the BLAS interface:
\listing{Grid/algorithms/deflation/MultiRHSDeflation.h}
This routine is more generally useful than multigrid, and also optimises regular fine
grid deflation calculation by an order of magnitude for several right hand sides.

\section{Results}

\label{sec:results}

We consider two test systems using the Mobius fermion
action with scale factor $b+c=2.0$ and $b-c=1.0$. These
have a physical point input quark mass $m_q=0.00078$ and
$L_s=24$. The $48^3\times 96$ ensemble is described in 
ref.~\cite{RBC:2014ntl}. The $96^3\times 192$ ensemble is not yet published but has been generated to support the constraint of finite volume effects
in RBC-UKQCD's effort to determine the hadronic vacuum polarisation contribution to muon g-2\cite{RBC:2023pvn}.

We first analysed the spectrum of the fine grid operator
on the $48^3$ volume, where we have determined the upper eigenvalue (power method) and 200 low lying eigenvalues using the Chebyshev preconditioned implicitly restarted Lanczos algorithm.
These are $\lambda_{\rm max} = 89.757$ and
$\lambda_{\rm min} = 6.92238\times10^{-6}$ giving condition number $\kappa = 12.88\times 10^6$. The low
lying portions fine and coarse operator 
 spectrum are displayed in figure~\ref{fig:exterior}.
The Chebyshev worst case bound\cite{saad} on convergence rate per iteration is thus, \beq\sigma = \frac{\sqrt{\kappa}-1}{\sqrt{\kappa}+1} = 0.999445. \label{eq:bound}\eeq

\subsection{Conjugate gradient baseline on test systems}

\begin{figure}
    \centering
    \includegraphics[width=0.6\linewidth]{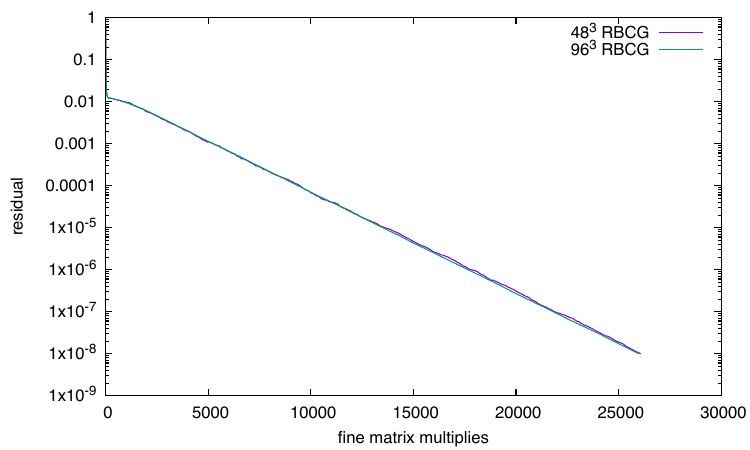}
    \caption{Convergence of the red-black preconditioned conjugate gradient algorithm with physical
    quark mass on the Mobius domain wall fermion action with the Schur operator ${\cal H}$ and on our
    $1.7$ GeV $48^3\times 96 \times 24$ (5.5fm) and  $96^3\times 192 \times 24$  (11fm) test configurations.
    The volume of eigenvector data and memory required to deflated the latter configuration is prohibitive as
    the cost grows as the square of the volume, whereas the multigrid solver setup, cost and footprint is proportional
    to the volume. The convergence is dictated by the spectrum and since the lattice spacing, quark mass and fermion action are identical between the two
    volumes (which differ by a factor of 16) the spectral density and convergence history remain almost identical. The system is solved with a polynomial
    order that is far less than the rank of both systems, which have a dense spectrum, so that each crossing of residual polynomial is covering a large number of eigenvalues in both cases, meaning the
    worst case bound is a reasonable description. }
    \label{fig:convergence-cg}
\end{figure}

Figure~\ref{fig:convergence-cg} displays the
convergence history of both the original red-black
preconditioned conjugate gradient for the Schur
complement operator ${\cal H}$, Eq.~\ref{Eq:MpcDagMpc}.
If we fit the long tail in residual history, from 5000 to 25800 iterations, to the form $r_n=A \sigma_{cg}^n$, we obtain a mean convergence rate:
\beq
\sigma_{cg} = .999451, \label{eq:convergencerate_fit}
\eeq
showing that the Chebyshev worst case bound Eq.~\ref{eq:bound} is saturated in our systems to a quite impressive accuracy. 
There is no statistical error on these figures since this is a pure algorithmic study for which single configuration analysis limits
cost and achieves adequate algorithmic conclusions. Thus the comparison between these figures is more than adequate to conclude practical consistency.
This is likely a reflection of the fact that our matrices are very large 
to the extent that the spectrum is in practice dense and reflects the limit where it is in practice
continuous compared to the CG iteration counts. There is little room for the optimal polynomial to
tune for individual eigenvalues because the order of the polynomial is still massively below
the order of the system. Each polynomial `ripple' around the true solution is still having
to accommodate thousands of eigenvectors leading to the near saturation of the worst case minimax bound.

This level of agreement suggests a simple method to estimate
the spectral range of the Dirac operator. 
The while the highest eigenvalue of the Dirac operator
is cheap to estimate using the power method, the lowest eigenvalue
requires an expensive iterative eigensolver.
It appears that the CG convergence rate in the long tail
of the convergence history, which is often available as part of a gauge configuration generation algorithms, gives an adequate estimator
for tasks such as setting a conservative spectral radius bound
for rational approximation (such as enters the RHMC algorithm\cite{Clark:2006fx}) and potentially simplifies the process of monitoring to ensure that spectral bounds set in parameters are preserved during RHMC evolution.

\clearpage 

\subsection{Basis completeness}

\begin{figure}
    \centering
    \includegraphics[width=0.5\linewidth]{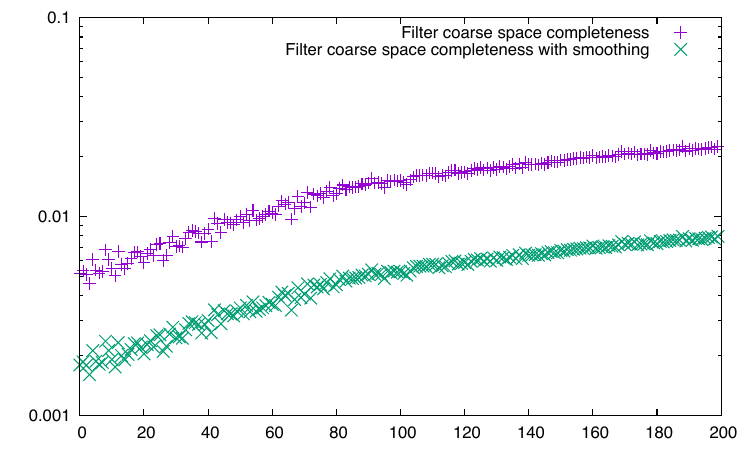}
    \includegraphics[width=0.49\linewidth]{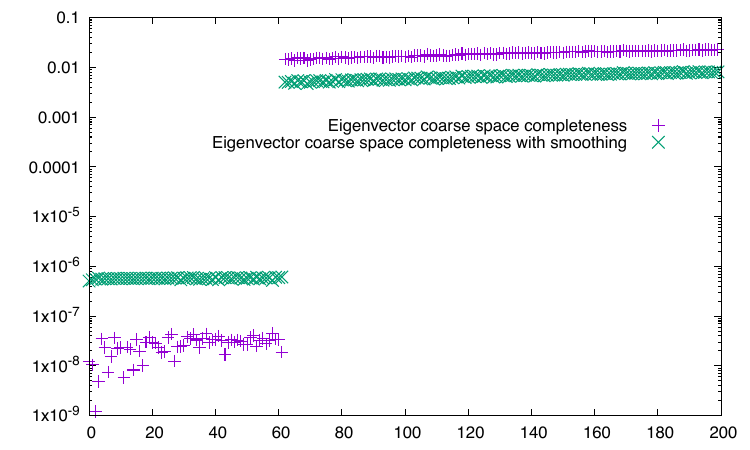}
    \caption{On our $48^3\times 96$ test volume we can assess the completeness of the 
    low mode by computing $
E_i = \sqrt{|| (1 - PP^\dagger) | i \rangle ||}$
and 
$E_i^{smoothed} = \sqrt{|| M_{IRS}(1 - PP^\dagger) | i \rangle ||}$ for each eigenpair, here with the Filter based subspace setup (left) and Lanczos based setup (right). Post-smoothing
reduces the error and may be indicative of a relative cheap way to improve coarse operator
based low mode variance reduction.
The exact eigenvectors included in the coarse basis (right) are faithfully reproduced to numerical precision, while higher modes have a percent scale error.}

    \label{fig:completenessFilter}
    \label{fig:completenessLanczos}
\end{figure}

The degree to which local coherence, or the weak approximation property, is numerically satisfied
by eigenvectors in QCD Dirac matrices can be quantified if a sizeable set of the the lowest modes are
computed using the Lanczos or other algorithms.
The quality of a restriction operator can be assessed by how well these low eigenpairs $\lambda_i, |i\rangle$ are preserved through the process of restriction and prolongation with the fractional error given by, for example,
\cite{Clark:2017wom,Brower:2018ymy},
$$
E_i = \sqrt{|| (1 - PP^\dagger) | i \rangle ||}.
$$
We compare the completeness of our coarse grid restriction and promotion with either Filter based
set up, figure~\ref{fig:completenessFilter} and Lanczos eigenvector based set up, fig:~\ref{fig:completenessLanczos}.

This is also performed with and without post-smoothing of the promoted result vector. 
$$
E_i^{smoothed} = \sqrt{|| M_{IRS}(1 - PP^\dagger) | i \rangle ||}.
$$

The use of a post-smoothing
step is included to to give some indication of the possible gain in precision to be had if using the coarse
grid inverse itself in low mode averaging variance reduction techniques, and also to illustrate the
potential for reduction of high mode spectral leakage back into the problem if smoothed aggregates\cite{alphaSA} were used in place of our disjoint blocks. Given the gain from smoothing is bounded by an order of magnitude
and the smoother was highly non-local, the direction of smoothed aggregation is less encouraging than
the vast increase in precision obtained in ref~\cite{Clark:2017wom} by using a larger basis and smaller blocking
factors.
Both schemes reproduce the low modes at the percent scale when the basis size and blocking are identical,
with exact reproduction of the lowest $n_{basis}$ modes with the Lanzcos based setup, while the modes
not included in the coarsening basis are actually slightly better reproduced in the filtering setup scheme.

\clearpage 
\subsection{Coarse operator spectrum}

\begin{figure}
    \centering
    \includegraphics[width=0.5\linewidth]{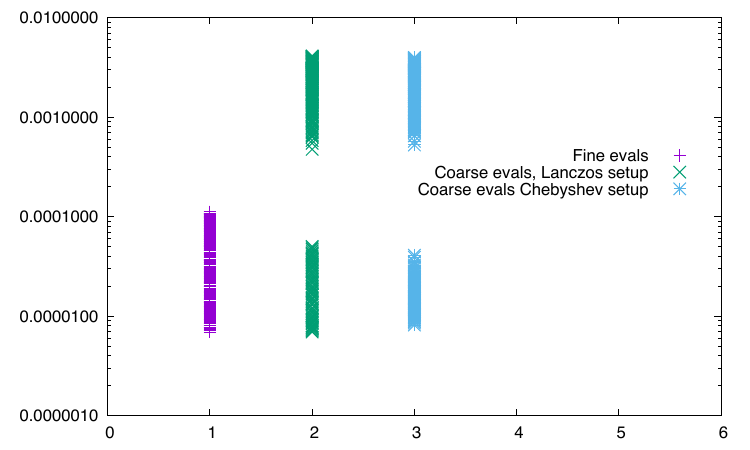}
    \includegraphics[width=0.49\linewidth]{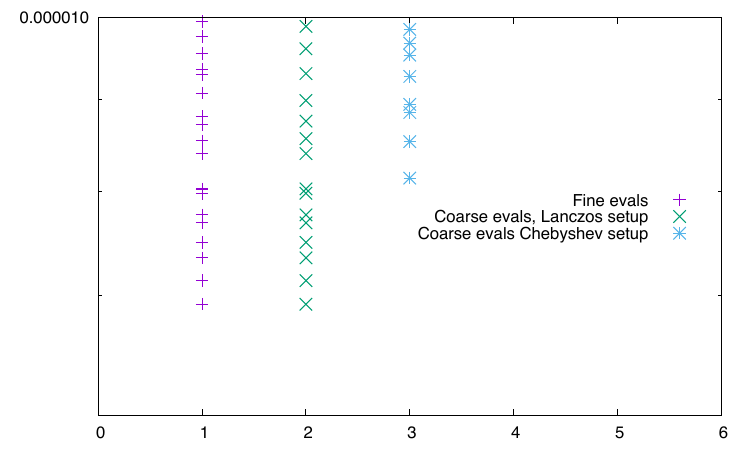}
    \caption{Spectrum of the lowest 200 eigenmodes of the fine operator ${\cal H}$ and the coarse
    operator with both Lanczos (green) and Chebyshev filter (blue) setup schemes with $62$ near null basis
    vectors on the $48^3$ lattice.
     A cluster of exactly $n_{basis}=62 $ very low eigenvalues are seen in the coarse operator, corresponding
     to the maximal diagonalisation of the fine operator within this set of near null vectors, whereas directions
     that involve a non-trivial coarse coordinate dependence in the coarse eigenvector necessarily incur spectral leakage
     at the boundaries between blocks and this lifts the coarse eigenvalue by an order of magnitude.
     The upper eigenvalue of the fine operator is around 89.0 while the upper eigenvalue of the coarse operator is around 37.0.
     On the right panel we zoom in on the lowest eigenvalues. We see that with the
    eigenvector setup the lowest eigenvalues are exactly reproduced by the coarse operator. This likely contributes  a to a slightly improved convergence rate in section~\ref{sec:solverresults}.
     }
    \label{fig:exterior}
    \label{fig:exterior-zoom}
\end{figure}

We can compare the eigenvalues determined for the coarse operator to those for the fine operator,
with the two subspace setup schemes figure~\ref{fig:exterior}.
The first feature with recognising is that with either setup scheme, given $n_{basis}$ near null vectors to define the coarse space there is necessarily
a subspace of the coarse linear space that involves globally rotating
within this space and without imposing coefficients that depend on the coarse
coordinate. This subspace necessarily continues to lie within
the near null space of the fine operator, and gives rise to a detached
cluster of exactly $n_{basis}$ near null eigenpairs for the coarse
operator, whose eigenvalues are close to the lowest eigenvalues of the fine
operator.
There is a second sub-space within the coarse representation that
gives rise to a bulk spectrum with lifted eigenvalues, owing to the
fact that directions in the coarse space that have coefficients
that depend on the coarse coordinate necessarily apply a `Lego-land' approximation to the low modes and the algebraic discontinuities between
our disjoint set of wavelets incur spectral leakage at the joins, lifting
the coarse eigenvalues by as much as an order of magnitude.

We zoom into the lowest modes in figure~\ref{fig:exterior-zoom}, 
and see that when the lowest $n_{basis}$ eigenvectors are used to
define the coarsening, both these vectors and their eigenvalues are faithfully represented in the coarse operator, while the eigenvalues
are only approximately represented in the Filter based setup scheme.

We will show that this has a marginal impact the the efficacy of the coarse
grid deflation, but likely requires that we reapply the preconditioner
sufficiently frequently that the coarse grid deflation is reapplied every time the residual in the outer iteration has been reduced to this level of approximation. In usage the
coarse Grid solver in the preconditioner is only solved to a tolerance of $5\times 10^{-2}$ in each outer iteration.

\begin{figure}
    \centering
    \includegraphics[width=0.5\linewidth]{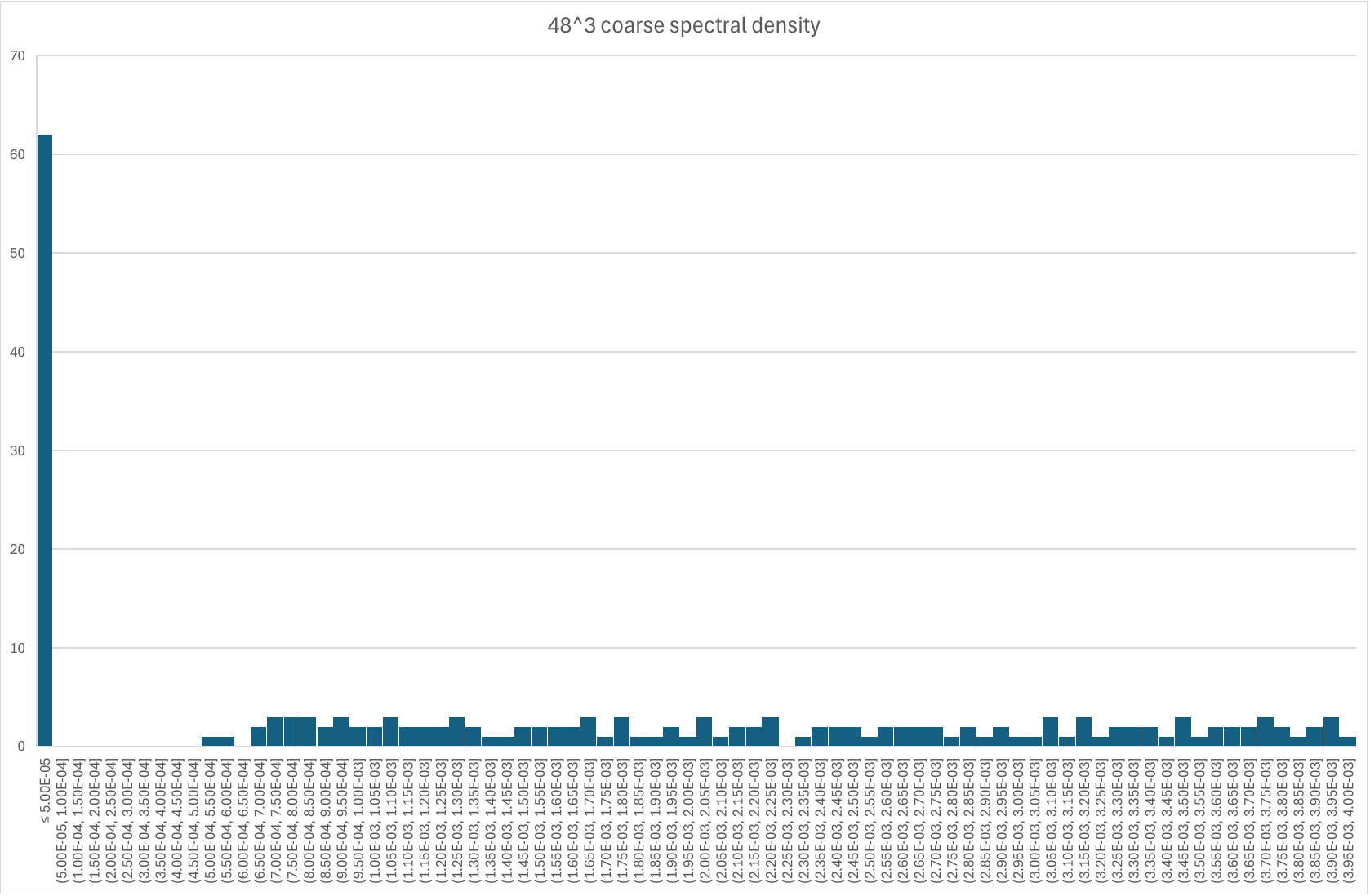}
        \includegraphics[width=0.49\linewidth]{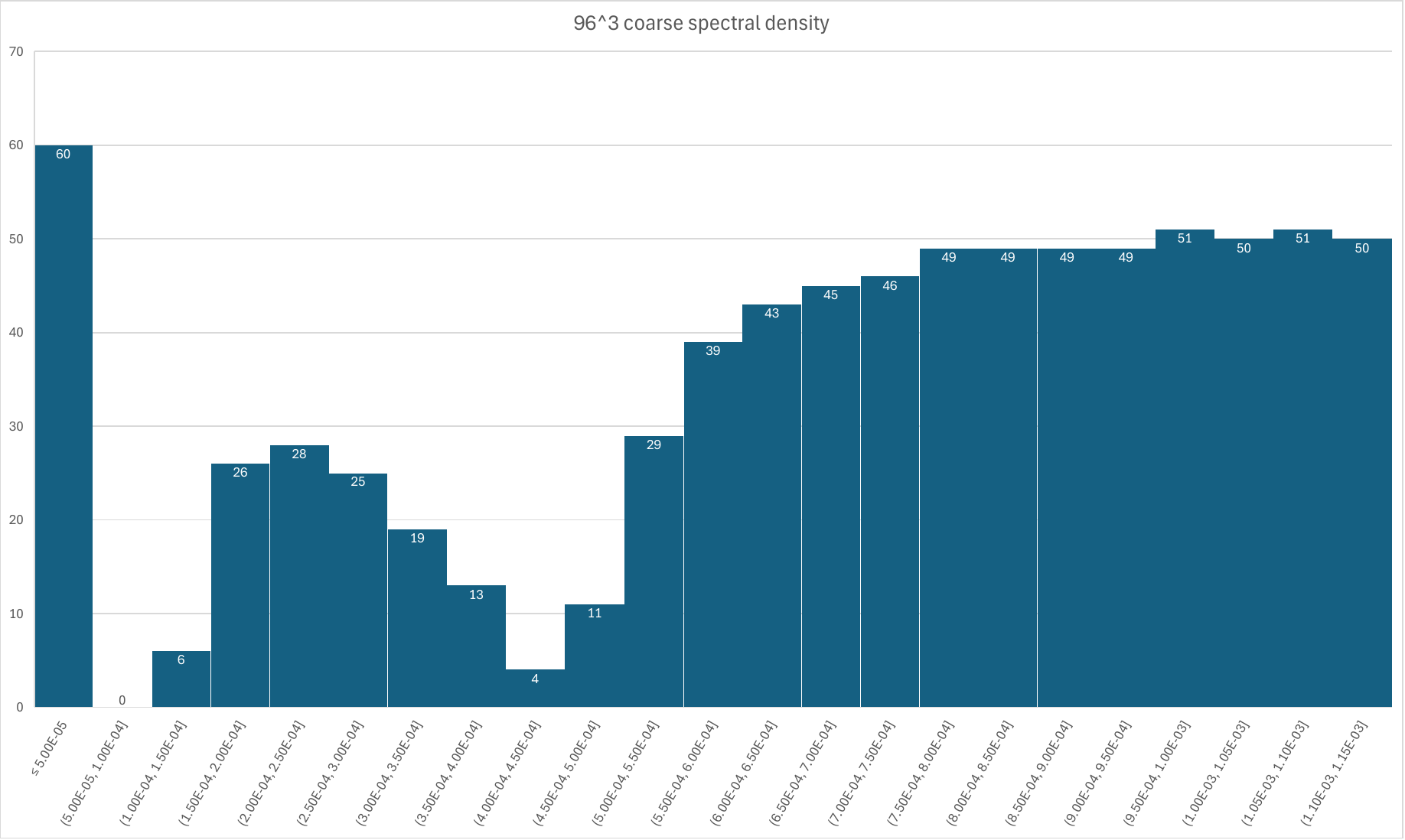}
    \caption{Spectral density (modes per bin, bin width $5\times 10^5$)
    for the coarse operator on the $48^3$ volume (left) and $96^3$ volume (right). There is a peak of $n_{basis}=62$ modes in the correct physical low mode region, corresponding to the dimension of the input set of near null vectors,
    with the eigenvectors in practice found by diagonalising mainly
    within this basis. Directions outside this sub-space necessarily induce
    upwards spectral leakage leaving this cluster clearly detached
    from the bulk spectrum with about 2 or 3 modes per bin.
    If we compare these, we see that the
    detached cluster again corresponds to the number of basis vectors defining the coarsening, but the density of modes in the higher, bulk spectrum
    grows linearly in the volume by exactly the expected factor of 16.
    }
    \label{fig:48dnsty}
\end{figure}

In somewhat more detail, the spectral density of the coarse
operator is plotted via a fixed bin-width histogram for both the
$48^3$  and $96^3$ 
volumes (figure~\ref{fig:48dnsty}). The number of modes in the detached low mode cluster is
given by the dimension of the near null space setup-vectors, 
whereas the more approximate bulk spectrum has  eigenvalues raised by
an order of magnitude due to spectral leakage, and a density that is growing
linearly in the volume as one expects.

While we used the Block Lanczos algorithm to compute eigenpairs
which are used to deflate the low mode space, given the detached
clustering it is possible that a faster setup option that diagonalises within
only the original set of near null space vectors is sufficient
to deflate the coarse spaces to a good degree, particularly
if looking for a fast setup algorithm for use in hybrid Monte Carlo.

\subsection{Solver performance}

\label{sec:solverresults}

\begin{table}[]
    \centering
    {\small 
    \begin{tabular}{|c|c|c|c|}
    \hline
         & $48^3\times 96$ & $48^3\times 96$ & $96^3 \times 192$\\
        \hline
    Algorithm & Flex-ADEF2& PrecBlockCGrQ & Flex-ADEF2\\
         $n_{basis}$& 62 &  62,64 & 60\\
         $n_{rhs}$& 12 &  12 & 12\\
         block& $4.4.6.4$ &  $4.4.6.4$ & $4^4$  \\
         Smoother & $M_{IRS}(2.0,7)$ & $\mathbb{P}^{cheby}(2.0,92.0,8)$  &$M_{IRS}(2.0,7)$\\
         Coarse solve & Defl CG tol $0.04$ & $\mathbb{P}^{cheby+defl}(0.02,40,120)$  &Defl CG tol $0.04$\\
         Coarse EVs & 192 & 192 & 1000\\
         Setup & Lanczos/Cheby & Lanczos & Cheby \\
         \hline
         Speedup & 12x - 15x & 20x & 10.5x\\
         \hline
    \end{tabular}
    }
    \caption{Parameters used to construct the multi-rhs multigrid on the different test cases and volumes. The key outcome of this paper is that all these multi-RHS algorithms give a significant
    gain exceeding and order of magnitude on modern GPU computing nodes, and that the fastest
    is the use of preconditioned BlockCGrQ, despite the requirement to freeze the multigrid
    preconditioner as a static polynomial (plus complementary deflation on the coarse space).
    The Block algorithm reduces outer iteration count by around 25\% and the static
    polynomial based multigrid preconditioner also executes faster than a Krylov based
    preconditioner due to the elimination of linear algebra and reductions in particular.}
    \label{tab:params}
\end{table}

In table~\ref{tab:params} we display the parameters used during our optimisation of the multi-right-hand side multigrid.
After building the subspaces as described in section~\ref{sec:subspace} the coarse
operators were computed and low lying eigenpairs are precalculated and stored using block Lanczos\cite{Jang:2018cph}.

 \subsubsection{Filter setup}

\begin{table}[htbp]
  \centering
    \begin{tabular}{|p{2.5cm}|l|l|p{1.5cm}|p{2cm}|p{1cm}|r|}
    \hline
    Setup & $n_{basis}$&Algorithm & Outer Iters & Fine matrix multiplies & 12 solves (s) & Speed up \\
    \hline
    $48^3\times 96\times 24$ &&&&&&\\
    \hline
    EV &64 & PrecBlockCGrQ  & 108   & 1080  & 460   & 20.2 \\
    EV &62& PrecBlockCGrQ  & 113   & 1130  & 480   & 19.3 \\
    EV &62 & FlexADEF2$^\ast$ & 131   & 1300  & 570   & 16.3 \\
    EV &62 & FlexADEF2 & 141   & 1269  & 602   & 15.4 \\
    Filter &62 & PrecBlockCGrQ & 123   & 1230  &  499  & 18.6 \\
    Filter &62 & FlexADEF2 & 167   & 1503  & 720   & 12.9 \\
    EV (sum-2) & 62 & FlexADEF2 & 213   & 1917  & 934   & 9.9 \\
    EV ($i|2=0$) & 62 & FlexADEF2 & 255   & 2295  & 1090  & 8.5 \\
    EV ($i|3=0$) & 62 & FlexADEF2 &268   & 2412  & 1182  & 7.8 \\
    \hline
    - & - & RedBlackCG & 26075 & 26075 & 9288  & 1  \\
    \hline
    \end{tabular}
  \caption{
       \label{tab:data48}
Solver performance of several optimised algorithms within the
  multi-RHS HDCG class/parameter space on the $48^3\times 96\times 24$ test system. The setup is either using Lanczos derived eigenvectors
  or Chebyshev based filters to determine the coarse space projectors, 
  with either 62 or 64 vectors. The outer solver algorithm is either a) PrecBlockCGrQ, our preconditioned block conjugate gradient with a static Chebyshev
 based smoother and deflated coarse grid correction as the multigrid preconditioner, or b) FlexADEF2 the flexible ADEF2 algorithm with a CG based smoother and deflated coarse grid correction as already described, or c) FlexADEF2$^\ast$ which combines our static multigrid
preconditioner with the FlexADEF2 algorithm. 
 The best algorithmic and floating point performance is obtained with our
 Preconditioned BlockCGrQ algorithm, yielding an overall speed up of
 20.3x gain per right hand side. We
 compared using the lowest $n_{basis}$ eigenvectors in our subspace
 setup, using the sum of consecutive eigenpairs and using every second
 and every third eigenvector. The most favourable spectral content / spectral distribution entering the multigrid coarsening appears very clearly to be representing purely the lowest $n_{basis}$ modes.
 While this could be a volume dependent conclusion, the $48^3$ volume.
These times were obtained using 18 nodes of the Frontier supercomputer at
Oak Ridge National Laboratory under the INCITE program.
 }
\end{table}

\begin{table}[htbp]
  \centering
    \begin{tabular}{|p{2.5cm}|l|l|p{1.5cm}|p{2cm}|r|r|}
    \hline
    Setup & $n_{basis}$&Algorithm & Outer Iters & Fine matrix multiplies & 12 solves (s) & Speed up \\
    \hline
    $96^3\times 192\times 24$ &&&&&&\\
    \hline
    Filter &60 & FlexADEF2 & 187  &  1683 & 1060  & 10.5 \\
    \hline
    - & - & RedBlackCG & 25984 & 25984 & 12720  & 1  \\
    \hline
    \end{tabular}
  \caption{
       \label{tab:data96}
Solver performance of several optimised algorithms within the
  multi-RHS HDCG class/parameter space on the $96^3\times 192\times 24$ test system. The setup uses only 60 Chebyshev based filters to determine the coarse space projectors.
  The outer solver algorithm is either a) preconditioned block conjugate gradient with a static Chebyshev
  based smoother and static Chebyshev based deflated coarse grid correction as the multigrid preconditioner, or b) the flexible ADEF2 algorithm with a CG based smoother and deflated CG based coarse grid correction as described in section~\ref{sec:static}. 
These times were obtained using 216 nodes of the Frontier supercomputer at
Oak Ridge National Laboratory under the INCITE program.
 }
\end{table}

\begin{figure}
    \centering
    \includegraphics[width=0.5\linewidth]{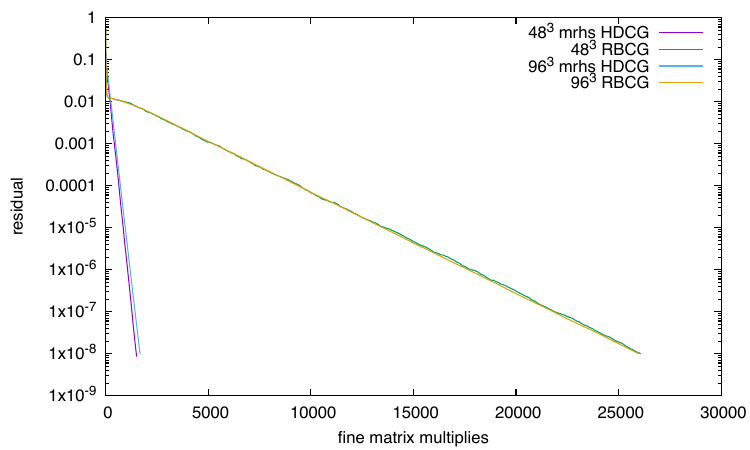}
    \includegraphics[width=0.49\linewidth]{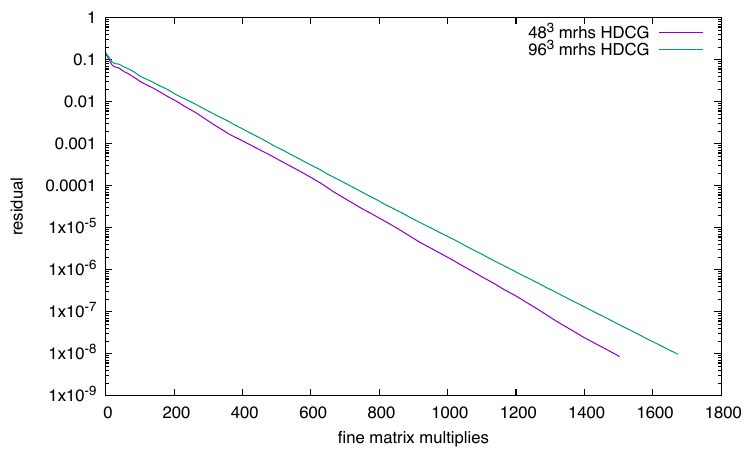}
    \caption{(Left) Convergence of mrhs-HDCG and red-black preconditioned Conjugate Gradient on sample $48^3\times 96$ and $96^3\times 192$ configurations. (Right) Zoomed comparison between the two volumes on the
    HDCG convergence history. 
    } 
    \label{fig:convergenceall}
\end{figure}

Figure~\ref{fig:convergenceall} left displays the
convergence history of both the original red-black
preconditioned conjugate gradient for the Schur
complement operator $M_{pc}$, and the
convergance history of the mrhs-HDCG (multiple right
hand side heirarchically deflated conjugate gradient). A factor of seventeen reduction in fine
matrix multiplies per right hand side solved is achieved.
Figure~\ref{fig:convergenceall}  right zooms in and displays only the convergence of the mrhs-HDCG
on both $48^3$ (with 62 Chebyshev filtered subspace vectors), and $96^3$ with 60 Chebyshev filtered 
subspace vectors. The convergence rate differs very little between the volumes and the small
difference is largely attributable to the slightly larger subspace on $48^3$ giving a little better
deflation fidelity. Deflation efficacy is broadly similar between the two volumes, while the only part of the valence solver algorithm
    that is proportional to $O({\rm volume})^2$ is the negligble coarse
    grid deflation, at less than 0.05\% of the run time.
    We can reasonably claim to have eliminated $O({\rm volume})^2$ for the
    forseeable future.

We now refer to exact iteration counts and execution times
listed in Table~\ref{tab:data48} for the $48^3$ ensemble using 18 nodes of the Frontier
supercomputer at Oak Ridge National Laboratory, and to  Table~\ref{tab:data96}
for the $96^3$ ensemble using 216 nodes of on Frontier.

The gain in matrix multiplies is from 26075 (CG) to
1503 for the filter setup on $48^3$, where the filter based setup FlexADEF2 algorithm takes 167 outer iterations. This invoolves with two residual computations and seven CG smoother iterations for a total of nine matrix multiplies per outer iteration. 
The gain in execution time per solve is a factor of 12.9 for the $48^3$ volume. 

For the $96^3\times 192$ volume, we had to reduce the multigrid blocking factor somewhat to retain the same deflation efficacy, blocking by $4^4$ rather than $4^3\times 6$, presumably by providing better eigenvector fidelity in the coarse
space compression.
The gain in execution time per solve is a factor
of 10.5 for the $96^3$ volume, reflecting two effects.
Firstly we took $n_{basis}=60$ subspace vectors in the setup rather 
62, and secondly the smaller blocking factor led to a larger
coarse volume and additional cost in the coarse space but was overall
the optimal choice on $96^3$.

The maximum possible gain from HDCG with this coarsening is therefore around seventeen fold, whereas for the total time to solution, additional overhead from the coarse grid correction and a general loss of efficiency in very short CG sequences in the smoother reduces the gain to around 12.9x and 10.5x on $48^3$ and $96^3$ volumes respectively for the filter based
subspaces and FlexADEF2.

For the HDCG convergence history (measured in outer steps), the convergence rate is 
\beq
\sigma_{hdcg} = 0.90693,
\eeq
and this allows an estimate of the condition number of the preconditioned system of $\kappa_{precon}\sim 419.8$.
We can assume that the upper eigenvalue of the preconditioned system is set by the low pass parameter
in the smoother, $\Lambda=2.0$, since this is the Dirac eigenvalue below which the high mode preconditioner
becomes ineffective. Using this assumption and the inferred outer condition number it appears that the 
composite preconditioner is also effective for Dirac operator eigenvalues \emph{below} $\lambda \sim .00476$.
This indeed matches plausibly well the upper end of the cluster of eigenvalues found by Lanczos for our course operator,
certainly extends almost two orders of magnitude above the range eigenvalues found by the fine grid Lanczos on the first
200 eigenpairs. There is therefore reasonable evidence that the multigrid preconditioner is effectively capturing a very considerable
number of low modes of the fine operator. 

\subsubsection{Eigenvector based setup}
 
We also used exact Lanczos derived eigenvectors used to define the coarse basis, on the $48^3$ volume. We obtained slightly better
performance than our filtered vector scheme, albeit at the expense of  about a factor of four slower multigrid setup cost. The outer iteration count decreased to 141 and the speedup increased to a factor of 15.4 gain over red-black preconditioned conjugate gradients.

When we introduce the fixed Chebyshev multigrid preconditioner the iteration count reduces
further to 131, with a modest gain in time to solution (16.3x faster). 
The introduction of Preconditioned BlockCGrQ with our $n_{basis}=62$ Lanczos derived eigenvectors improves
the outer iteration count by around 25\% and the speed up increases to over 19x, while a psychologically
important threshold of 20x speed is obtained with $n_{basis}=64$.

There has been considerable speculation in the field, with relatively little hard information, that 
the inverse iteration commonly used in multigrid is more
effective because it produces spectrally broad vectors
overlapping with a wide range of eigenvectors in the near null space, rather than project purely on the very lowest subset of modes.

We therefore compared using every second and every third mode in addition to using purely the lowest $n_{basis}$ modes. We also compared using
linear combinations of consecutive modes. In these additional
cases, the performance was significantly worse than both our filtering
and Lanczos setup schemes. 

This leads to the following finding:
\begin{center}
\fbox{
\begin{minipage}{0.9\textwidth}
We find that the best set of multigrid deflation vectors arises by taking $n_{basis}$ pure eigenmodes in ascending eigenvalue order starting from the lowest eigenpair.
\end{minipage}
}    
\end{center}

These can either be obtained from the Lanczos algorithm, or if the cost is prohibitive the learning
is that projecting a linearly independent set of vectors that maximise the degree to which they are near null (nulliness) is the best filter setup strategy.

\subsubsection{Preconditioned BlockCGrQ}

Finally we consider the effect of using our Preconditioned BlockCGrQ algorithm with stationary preconditioner composed of a Chebyshev
smoother and deflated Chebyshev coarse grid inversion.
This reduces the outer iteration count from 169 iterations to 123 iterations in the Filter setup on the $48^3$ configuration, and the
outer iteration count from 141 iterations to 113 iterations with the Lanczos derived eigenvector setup with 62 basis vectors.
The difference between the more approximate filter coarse basis set up and the precise eigenvector coarse basis setup is
much reduced by the use of the Block solver algorithm, and suggests that the block solver is significantly more tolerant
of lower quality subspace creation.

\begin{figure}
    \centering
    \includegraphics[width=0.7\linewidth]{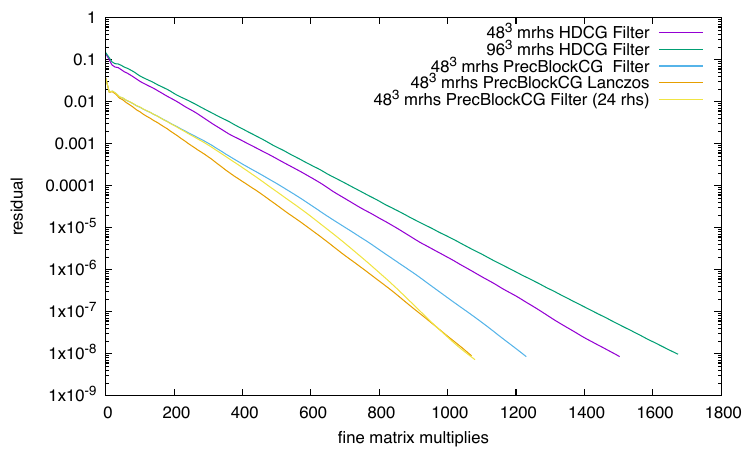}
    \caption{Convergence of Flexible ADEF2 and preconditioned BlockCGrQ on the $48^3$ configurations with 12 right hand sides and 62 basis vectors (either eigenvectors or filtered noise vectors). The
    Block algorithm substantially reduces the difference between the eigenvector and filtered vector basis creation choices.
    The Block algorithm appears to better tolerate an imperfect setup than preconditioned CG alone. 
    With 24 right hand sides and the Filter setup there is clearer evidence
    of the superlinear convergence property of BlockCG, however with the current software
    implementation the linear algebra overhead is too large to make this beneficial.
    } 
    \label{fig:convergenceblockcg}
\end{figure}

The convergence histories are shown in figure~\ref{fig:convergenceblockcg}
This approach leads to the overall fastest solver obtained within this paper, and a speed up of 22.5 over red black
preconditioned conjugate gradients on the $48^3$ ensemble. 
There is some evidence of superlinear convergence onset in the preconditioned BlockCGrQ history,
 with 12 right hand sides this is marginal curvature on the log plot, and this improves
 significantly with 24 right hand sides.

The breakdown of the runtime spent in different components of the optimised
solver is given in Table~\ref{tab:breakdown}. Here the only
element of the algorithm that is proportional to $(Volume)^2$ is the coarse
grid deflation. This is at that 1.5\% level of the total run time, and
is unlikely to become significant within a long period of time given the
computational cost of current and planned lattice calculations.
Therefore, it is reasonable to claim this work has completely
suppressed the  $(Volume)^2$ cost of deflation while obtaining a similar
algorithmic gain.

Since the number of basis vectors and the number of right hand sides are tunable parameters, it is
useful to illustrate that if we accept increased memory footprint, and possibly setup costs, we can
obtain further algorithmic gain in exchange for additional organisational overhead and memory footprint.
Extending to 24 right hand sides further reduces the iteration count from 123 iterations with the
filter setup to 108 iterations. The linear algebra in the Block solver
is proportional to the number of right hand sides squared, and this purely algorithmic reduction does not lead to an overall
faster solver due to the linear algebra increasing to over 20\% of the computer time.
It is possible this could be addressed with further optimisation.


\begin{table}[]
    \centering
    \begin{tabular}{|c|c|}
    \hline
     Operation & Time for component \\
     \hline
       {\color{red} Linear Algebra}  & 16s  \\
       Fine residual  & 30s   \\
       Multigrid preconditioner  & 368s  \\
       \hline
       Total  &  417s \\
       \hline
       {\color{red} Restriction}  & 3.5s  \\
       {\color{red} Prolongation}  & 2.9s  \\
       {\color{red} Coarse deflation}&   6s  \\
       {\color{red} Coarse solve } &  100s  \\
       Smoother &   223s  \\
       \hline
    \end{tabular}
    \caption{
    In the fastest Preconditioned BlockCGrQ run from the $48^3$
    ensemble we give the breakdown of the total time into the component operations.
    The operations that are accelerated using the (batched) GEMM interfaces are colored
    red, and consume around 30\% of the run time, illustrating how machine learning and artificial intelligence focussed matrix and tensor hardware can be used directly in multi-rhs multigrid solvers. Of this overhead, the coarse grid deflation is around 1.5\% but is the only element of the algorithm that scales as $O({\rm volume}^2)$
    }
    \label{tab:breakdown}
\end{table}

\subsubsection{Power spectrum analysis of error and residual}

\begin{figure}
    \centering
    \includegraphics[width=0.5\linewidth]{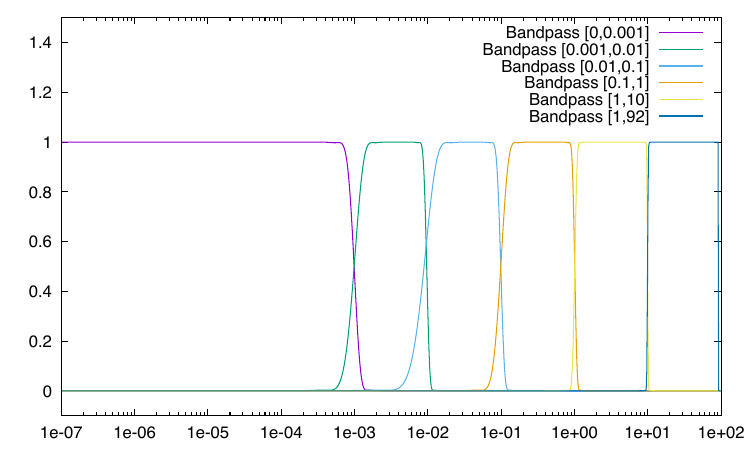}
    \caption{Chebyshev polynomial spectral bandpass filters used
    to evaluate the power spectrum of different operators entering
    our multigrid solver. This is an interesting probe and diagnostic
    of our algorithms.}
    \label{fig:JacksonThetaFunctions}
\end{figure}

We can use smoothed Chebyshev band pass filters, figure~\ref{fig:JacksonThetaFunctions}, of the squared red-black Dirac
  operator to measure the power spectrum, table~\ref{tab:PowerSpectrum} of the error and the residual in the HDCG solution vector during 
  convergence. We stopped the solver when the  when the residual is $10^{-3}$, $10^{-4}$ and $10^{-5}$ comparing
  the current solution to the ``true'' solution obtained with residual $10^{-8}$ in each case.
  The power is almost entirely localised in the lowest bin corresponding to eigenvalues between zero and $10^{-3}$.
  The total power measured using this method is 0.99943, so that only $6\times 10^{-4}$ of the norm of the error vector
  is unaccounted showing the method is quite precise.
  The  the power in the residual is almost entirely localised in the highest bin corresponding to eigenvalues between 10 and 100.
  These are amplified in the spectral content by the application of the matrix to the solution vector in forming $r=A x-b$.
  The total power measured using this method is 0.99761, so that only $3\times 10^{-3}$ of the norm of the residual vector
  is unaccounted showing the method is quite precise.

\begin{table}[htbp]
  \centering
    \begin{tabular}{|p{2cm}|p{1.9cm}|p{1.8cm}|p{1.8cm}|p{1.7cm}|p{2cm}|p{1.8cm}|}
  \hline
    Eigenvalue range & Solution Error  & Residual & $r$& $M_{IRS} r$ & $Q r$ &  $z=M^{-1} r$  \\
    \hline
    $[0,0.001]$ & 9.98E-01 & 3.95E-05 & 1.54E-06 & 3.71E-05 & 8.19E-01 & 5.08E-01 \\
    $[0.001,0.01]$ & 1.32E-03 & 1.59E-04 & 1.25E-05 & 3.01E-04 & 1.04E-01 & 6.81E-02 \\
    $[0.01,0.1]$ & 2.03E-04 & 2.55E-03 & 3.62E-04 & 8.57E-03 & 3.98E-02 & 3.84E-02 \\
    $[0.1,1]$ & 2.24E-05 & 2.14E-02 & 1.27E-02 & 2.50E-01 & 1.11E-02 & 1.15E-01 \\
    $[1,10]$ & 1.03E-06 & 5.35E-02 & 1.11E-01 & 6.19E-01 & 2.44E-03 & 2.13E-01 \\
    $[10,100]$ & 5.98E-08 & 9.20E-01 & 8.71E-01 & 9.36E-02 & 1.74E-03 & 3.06E-02 \\
    \hline
    Total & 0.99943 & 0.99761 & 0.99545 & 0.97072 & 0.97797 & 0.97413 \\
    \hline
    \end{tabular}%
  \label{tab:PowerSpectrum}%
  \caption{
    To illustrate the spectral behaviour of the multigrid preconditioner,
    we take the power spectrum of six different vectors with respect to the 
    red-black preconditioned Dirac operator.
    These vectors are (left to
    right)  the error on the solution vector when the solver is only converged to $10^{-4}$ accuracy and the residual at this $10^{-4}$ iteration. The error in the solution vector is predomninantly in the lowest bin, while the residual is dominated by the highest few bins.
    The final four vectors are the final residual, the smoother applied
    to this final residual (reduces power in the highest mode bin),
    the coarse grid inverse applied to the residual (dominated by lowest spectral bin)
    and the composite multigrid preconditioner applied to this residual, with quite
    an interesting spectral shape.
  }
\end{table}%

\subsection{Software Performance}

Considering algorithmic performance as somehow independent from computer performance
is a fallacy we must avoid. While simple Krylov space algorithms
can be loosely characterised by matrix multiplications, the cost
of linear algebra still varies. Once more than one type of operation
is included in our considerations the relative execution rate of these
operations is important to consider on real computers with constraints
such as memory bandwidth that extend beyond just counting floating point operations. Therefore the ``optimal'' algorithm or indeed algorithmic parameters on any given real world computer in fact depends on the characteristics of the system since the relative costs of different types of operation will change from system to system. This is an unfortunate reality, and it is important to assess the time spent executing the different operations used in our algorithms since only then can they be
translated into total execution time. It is for precisely this reason that a substantial effort has been 
invested in software optimisation in this work, including a wholesale use of batched BLAS libraries
and GPU tensor cores.

\begin{figure}
    \centering
    \includegraphics[width=0.9\linewidth]{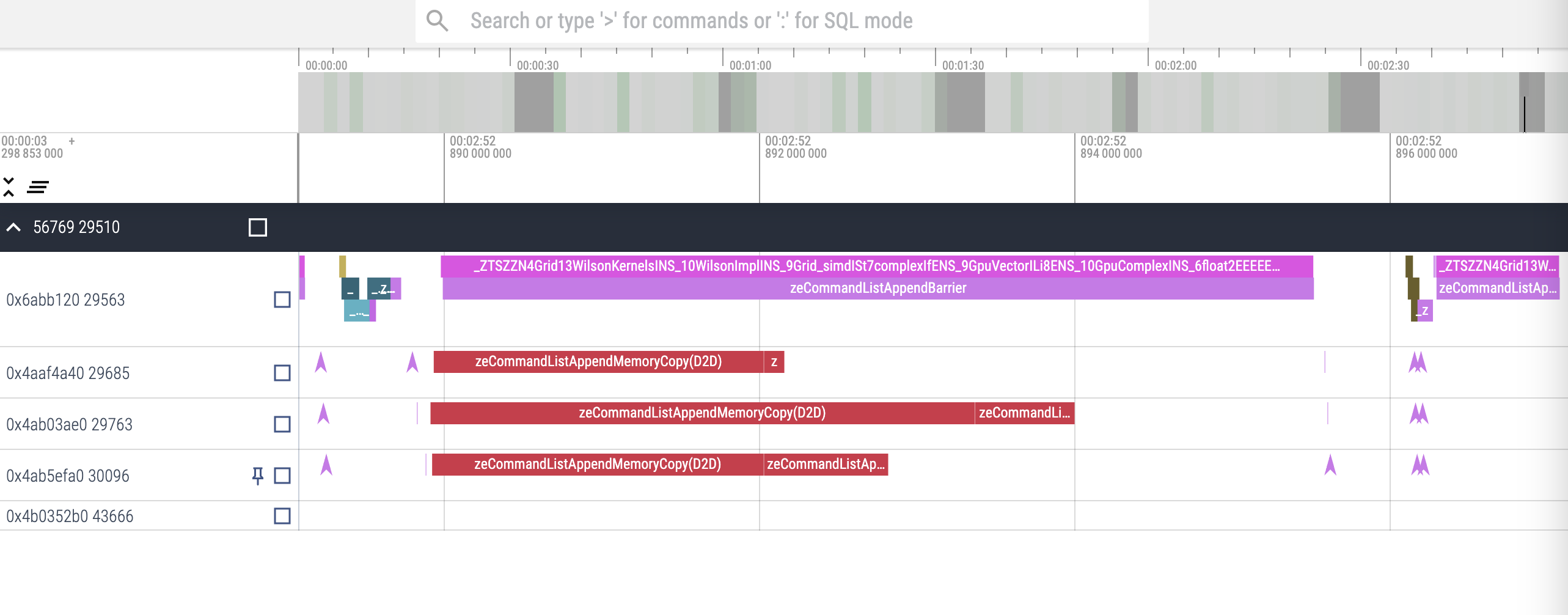}
    \caption{Profile of the fine grid domain wall operator running on a single node of the ANL Sunspot supercomputer. Intra-node communications are performed using DMA engines programmed via SYCL memcpy instructions and using the Level Zero API shared memory support. After gathering face data (multiple colours), communications (red) and computation (purple) are well overlapped, and the halo exchange    for the PDE stencil is performed concurrently with computation. The additional terms
    from the surface are added in as a second major kernel call.
    The performance in single precision is over 18 TF/s per node on large
    local volumes.}
    \label{fig:sunspot}
\end{figure}

The performance of Grid for the fine Grid five dimensional Wilson operator
that enters the domain wall formulation has been carefully optimised over many years
on all modern GPU systems\cite{Yamaguchi:2022feu}.
The software with a local volume of $32^3$ per GCD obtains 9TF/s per node
on Frontier comprising four AMD MI250X GPUs, 10TF/s on Tursa and Juelich Booster comprising
four Nvidia A100 GPUs, and 17TF/s on ANL Sunspot (noting the system is on early software and this may change
with later revision software) comprising six Intel Pontevecchio GPUs.
Communication and computation are overlapped very effectively.  Nvidia advisory profiling tool 
nsight compute indicates that memory bandwidth is fully saturated in each kernel.
An example profile from the Sunspot computer with a small local volume
is displayed in figure~\ref{fig:sunspot}.

\begin{figure}
    \centering
    \includegraphics[width=0.9\linewidth]{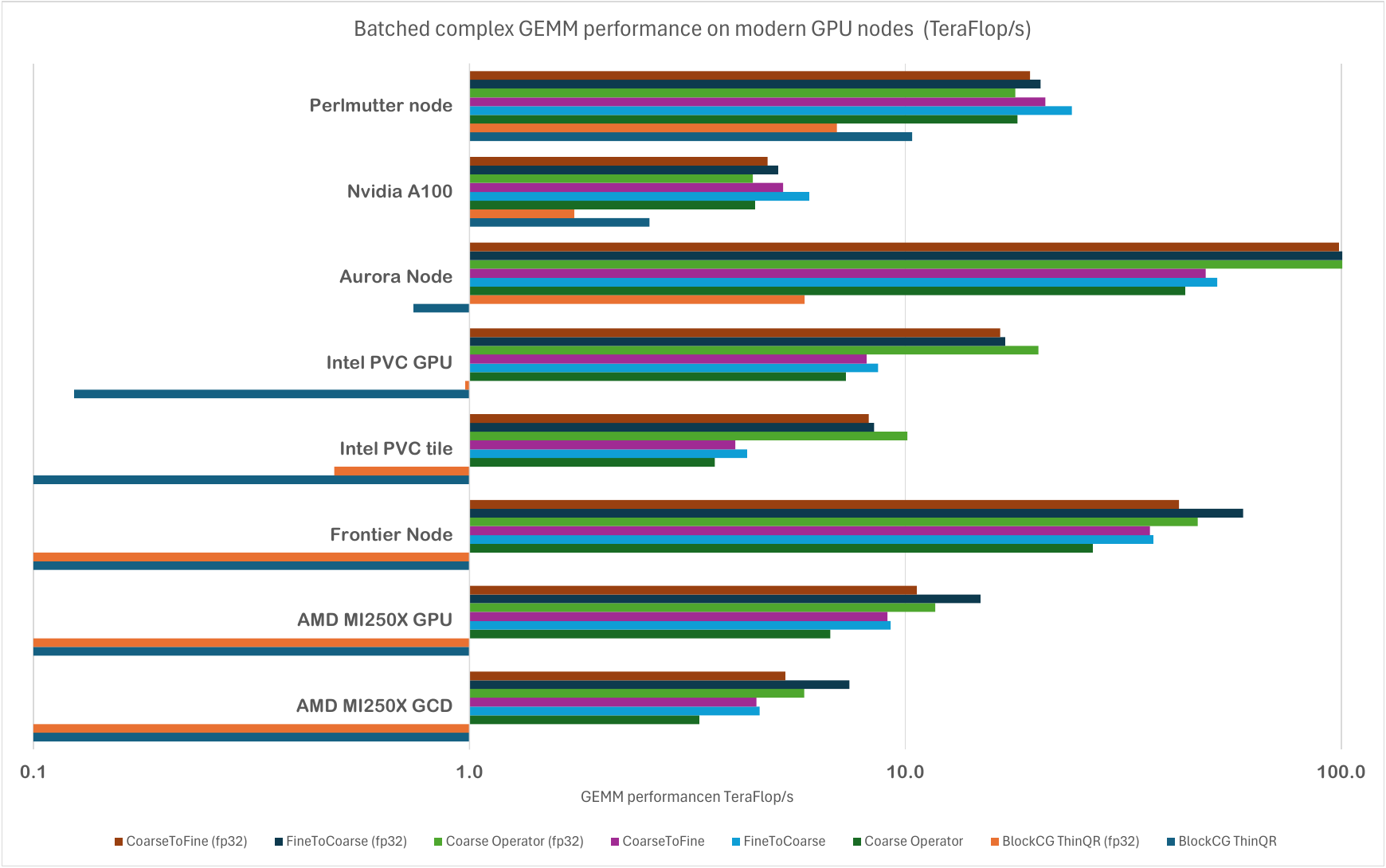}
    \caption{Performance of batched GEMM operations in TF/s as used in mrhs-HDCG across a variety of
  modern GPUs and some of the most
  significant current supercomputing platforms. 
  The Matrix dimensions correspond to the application of the coarse grid operator, the projection of data from the fine grid
  to the coarse grid, the promotion from the coarse grid to the fine grid
  and the QR rotation that enters the Block conjugate gradient algorithms.
  These include the Frontier supercomputer at ORNL 
  (four AMD MI250X GPUs
  and eight logical GCD's), the Sunspot/Aurora supercomputer at ANL (six Intel Pontevecchio GPUs and 12 logical tiles)
  and the Perlmutter supercomputer at NERSC. Multiple TF/s per logical GPU is easily obtained on most
  of the relevant matrix ranks, with the exception of the ThinQR factorisation in BlockCG on the fine grid on AMD and Intel libraries.
  This is relatively easily addressed in our implementation by using a batched call to execute many shorter $K$ matrix multiplications and then summing manually the resulting $12\times 12$ matrices, and then yields several
  TF/s per GPU, but the vendor library delivers less than 10 GF/s performance without this approach.}
    \label{fig:blas}
\end{figure}

The key addition to the software here is multigrid preconditioner. This has
been by design reduced to batched GEMM operations which are for the most
part highly optimised in vendor libraries, which we make use of.
In Table~\ref{tab:GEMMperf} we show the measured performance
(excluding communication) of the batched BLAS accelerated components of
our multigrid, and this is displayed graphically in figure~\ref{fig:blas}. Multiple TF/s per logical GPU is easily obtained on most
  of the relevant matrix ranks, with the exception of the ThinQR factorisation in BlockCG on the fine grid on AMD and Intel libraries.
  This is addressed in our implementation by using a batched call to execute many shorter $K$ matrix multiplications and then summing manually the resulting $12\times 12$ matrices, and then yields several
  TF/s per GPU. The optimised BlockCG linear algebra support is available on GitHub,

\listing{algorithms/deflation/MultiRHSBlockCGLinalg.h}

The complete outer iteration of our PrecBlockCGrQ solver is profiled in figure~\ref{fig:mgridprofile}, where the
smoother is a dense sequence of fine-grid matrix multiplies on 12 right hand sides, followed
by projection to coarse, deflation, coarse chebyshev solver, promotion to fine and then a thinQR 
rotation. This following sequence is almost entirely performed using batch BLAS opertions and
is displayed as such by the profiler with the blocks labelled as using matrix hardware.

\begin{table}[htbp]
  \centering{\small
    \begin{tabular}{c|ccc|c|ccc|ccc|cc}
        &         &        &       &  
        &\multicolumn{3}{c|}{Frontier}
       &\multicolumn{3}{c|}{Aurora}
       &\multicolumn{2}{c}{Perlmutter}\\
        &         &        &       &  
        &\multicolumn{3}{c|}{AMD MI250X }
       &\multicolumn{3}{c|}{Intel PVC }
       &\multicolumn{2}{c}{Nvidia A100 }\\
       \hline
          & M     & N     & K     & Batch & GCD &  GPU &  Node & Tile & GPU & Node  &  GPU &  Node \\
          \hline
          FP64 &&&&&&&& &&\\
    CoarseOp  & 64    & 12    & 64    & 4096  & 3.4   & 6.7   & 26.9  & 3.7   & 7.3   & 43.8  & 4.5   & 18.1 \\
    Project & 64    & 12    & 256   & 4096  & 4.6   & 9.3   & 37.1  & 4.3   & 8.7   & 52.0  & 6.0   & 24.1 \\
    Promote & 12    & 256   & 64    & 4096  & 4.6   & 9.1   & 36.4  & 4.1   & 8.1   & 48.9  & 5.2   & 21.0 \\
     ThinQR & 12    & 12    & 1M & 1     & 0.0   & 0.0   & 0.1   & 0.1   & 0.1   & 0.7   & 2.6   & 10.4 \\
     \hline
            FP32 &&&&&&&& &&\\
    CoarseOp  & 64    & 12    & 64    & 4096  & 5.9   & 11.7  & 46.9  & 10.1  & 20.2  & 121.2 & 4.5   & 17.9 \\
    Project  & 64    & 12    & 256   & 4096  & 7.4   & 14.9  & 59.5  & 8.5   & 17.0  & 101.7 & 5.1   & 20.4 \\
    Promote  & 12    & 256   & 64    & 4096  & 5.3   & 10.6  & 42.5  & 8.2   & 16.5  & 98.9  & 4.8   & 19.3 \\
     ThinQR  & 12    & 12    & 1M & 1     & 0.0   & 0.0   & 0.0   & 0.5   & 1.0   & 5.9   & 1.7   & 7.0 \\
    \end{tabular}%
    }
  \caption{Tabulated performance of batched GEMM operations as used in mrhs-HDCG across a variety of
  modern GPUs and some of the most
  significant current supercomputing platforms. 
  The Matrix dimensions correspond to the application of the coarse grid operator, the projection of data from the fine grid
  to the coarse grid, the promotion from the coarse grid to the fine grid
  and the QR rotation that enters the Block conjugate gradient algorithms.
  These systems include the Frontier supercomputer at ORNL 
  (four AMD MI250X GPUs
  and eight logical GCD's), the Sunspot/Aurora supercomputer at ANL (six Intel Pontevecchio GPUs and 12 logical tiles)
  and the Perlmutter supercomputer at NERSC. 
  }
  \label{tab:GEMMperf}%
\end{table}%

\begin{figure}
    \centering
    \includegraphics[width=0.9\linewidth]{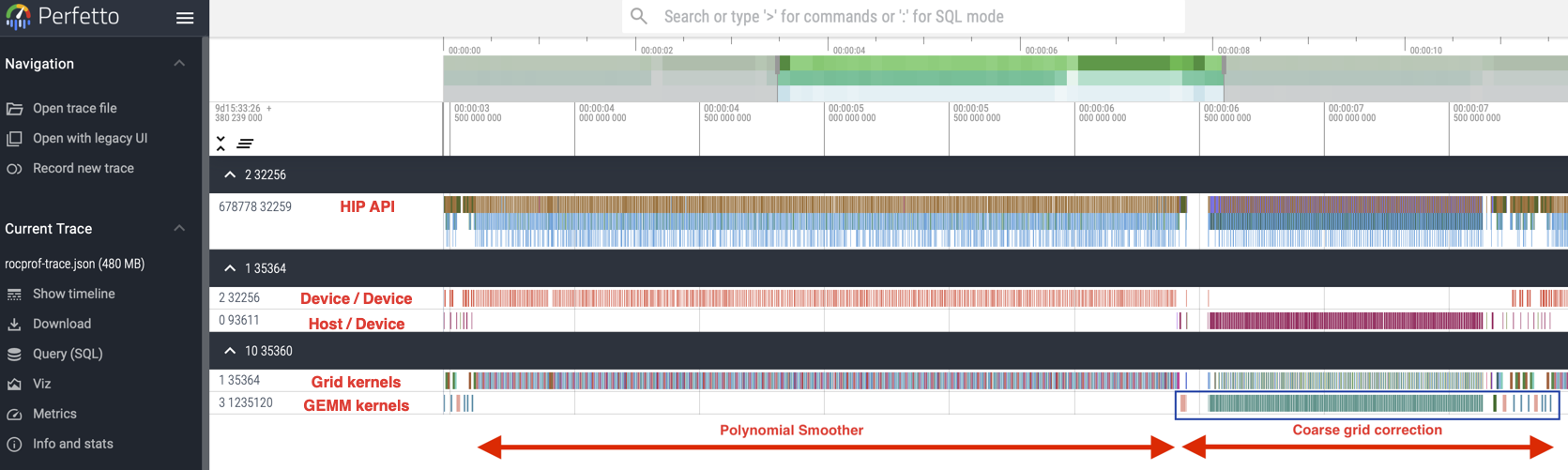}
    \caption{AMD Rocprof obtained profile from Frontier of the multigrid iteration on 18 nodes on the $48^3$ test problem after careful optimisation. The general Grid software kernels are
    shown executing along side GEMM kernels, used by projection to coarse, deflation, coarse Chebyshev solver, promotion to fine and then a thinQR 
    rotation. Broadly it is possible to perform almost the
    entire multigrid preconditioner in BLAS routines using
    optimised hardware,
    except for the relatively modest overhead of data layout changes and of course halo-exchange routines.}
    \label{fig:mgridprofile}
\end{figure}

\begin{figure}
    \centering
    \includegraphics[width=0.9\linewidth]{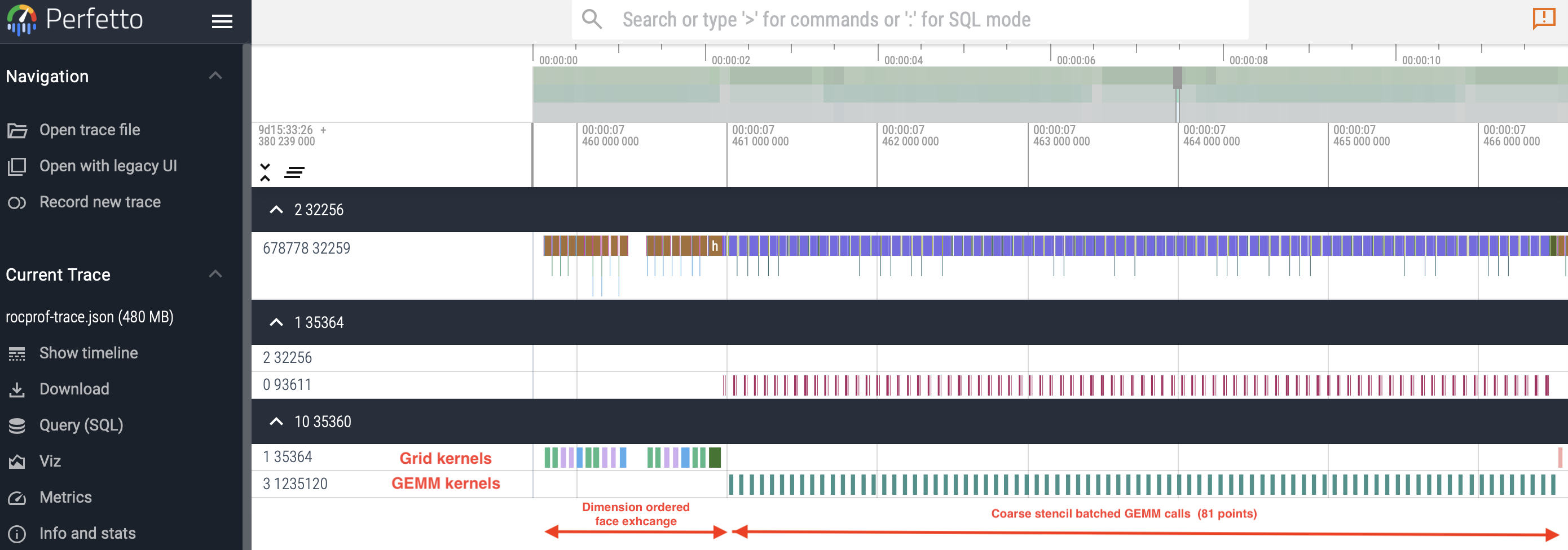}
    \caption{AMD Rocprof obtained profile from Frontier of the Coarse Grid operator on 18 nodes on the $48^3$ test problem  after careful optimisation. The general Grid software kernels are
    shown executing along side GEMM kernels. There remains some scope for further optimisation in the coarse space operator as GPU synchronisation overhead remains a 50\% overhead in that routine by fusing together larger batched operations, perhaps an estimated 10\% effect on the overall solver performance. }
    \label{fig:mgridprofile}
\end{figure}

\section{Conclusions and outlook}

The main findings of this paper are as follows.
We introduced a class of multiple right hand side multigrid algorithms for domain wall fermions that
speeds up valence propagator calculations at physical
quark masses by a factor of over twenty.

The solver algorithm is successfully designed to evade the $O({\rm volume}^2)$ cost of eigenvector deflation, and this is suppressed to the percent scale and arises only in the deflation of the coarse grid. This defers the need to further address $O({\rm volume}^2)$ for the foreseeable future.

We introduced a preconditioned BlockCGrQ algorithm which
can combine multigrid preconditioning with block algorithms
and leads to an additional 30\% speed up.
The algorithm is broadly applicable and may help
with multiple right hand sides multigrid in a broad range of contexts. The software techniques for optimising
the coarse multiple right hand side operations are also broadly applicable.

Using stationary Chebyshev preconditioners avoided the need for flexible algorithms and was computationally more efficient.
Were this not possible, we would have introduced a higher
degree of A-orthogonalisation of the Block search directions
with respect to recent history, as is common in flexible algorithms.

We found that multigrid subspace set up is optimal when obtaining a linearly independent set of vectors projecting as hard as possible on the lowest eigenvalue modes of the system. However the loss of convergence rate due to upwards spectral leakage is fairly slow.

There remains potential for substantial benefit from reducing the set up cost of multigrid algorithms, even for the Wilson fermion discretization, in order to fully realize the same benefit for gauge configuration generation
\cite{Francis:2019muy,Brower:1995vx}. Fast setup and
single right hand side optimisation
will remain the subject of future research.

The calculations in this paper have been entirely in double precision
to avoid mixing algorithmic with implementation details.
However, in ref.~\cite{Boyle:2014rwa} it was convincingly shown
that the smoothers and coarse solvers (which are only solved to low precision) can tolerate single precision execution and double precision
accuracy need only preserved in the outer solver. Thus around 90\% of the operations can be single precision accelerated.

Ref.~\cite{Boyle:2014rwa} also showed that the deflation impact is preserved under physically important perturbations of the matrix
being inverted, provided the coarse operator is recalculated, and without recalculation of the deflation basis vectors. 
The explored perturbations included changes to the quark mass and also twisted boundary conditions for the gauge fields.
We expect that this insight will also be preserved in the introduction of stochastic $U(1)$ gauge fields for modest coupling, appropriate
to calculation of the linear term in the QED coupling.

\section{Acknowledgements}

I wish to thank my colleagues in RBC-UKQCD and in the USQCD SciDAC-5 project for many useful conversations. PB has been supported in part by the Scientific Discovery through Advanced Computing (SciDAC) program LAB 22-2580, and by US DOE Contract DESC0012704(BNL).

Particular thanks to Christoph Lehner for motivating elements of this work by suggesting we port the HDCG algorithm to GPUs for the $96^4\times 192 \times 24$ volume problem. Christoph
also noted that eigenvector deflation would become intractable at these volumes. We thank Chulwoo Jung for developing and contributing the block Lanczos algorithm in Grid, and for providing advice on the use of the software.

The test runs performed in this paper
were carried out on the Frontier supercomputer
at Oak Ridge National Laboratory and on the
Summit supercomputer at Argonne National Laboratory.
Benchmarking was carried out on the DiRAC Tursa
computer at Edinburgh University and on the Perlmutter computer at NERSC.

\bibliographystyle{unsrt}
\bibliography{sample}

\end{document}